\documentclass[aoas,authoryear,preprint]{imsart}

\RequirePackage[OT1]{fontenc}
\RequirePackage{amsthm,amsmath}
\RequirePackage{natbib}
\RequirePackage[colorlinks,citecolor=blue,urlcolor=blue]{hyperref}

\startlocaldefs

\usepackage[english]{babel}
\usepackage[utf8]{inputenc} 

\usepackage[toc,page]{appendix}

\usepackage{mathrsfs,amssymb}  
\usepackage{theorem}	
\usepackage{accents} 
\usepackage{relsize} 

\usepackage{array}
\usepackage{tabu} 
\usepackage{booktabs} 
\usepackage{multirow} 
\usepackage{units}
\usepackage{threeparttable}

\usepackage{ifthen,calc}

\usepackage{fancyhdr}
\usepackage{parskip} 
\usepackage{nicefrac} 
\usepackage[super,negative]{nth} 
\usepackage{graphicx} 
\graphicspath{{./Figures/}} 
\usepackage{subfigure} 
\usepackage{float} 
\usepackage{wrapfig} 

\usepackage[textsize=footnotesize]{todonotes}

\newcommand{\bel}{\begin{equation}\label}
\newcommand{\be}{\begin{equation}}
\newcommand{\ee}{\end{equation}}

\DeclareMathOperator{\Var}{Var}

\renewcommand{\P}{\mathbb{P}} 
\newcommand{\E}{\mathbb{E}} 
\newcommand{\toop}{\stackrel{\P}{\longrightarrow}} 
\newcommand{\schw}{\stackrel{d}{\longrightarrow}} 
\newcommand{\Normal}{\mathcal{N}} 
\DeclareMathOperator{\Binom}{Binom} 

\newcommand{\kstar}{{k^*}} 

\newcommand{\Q}{T_{\chi^2}} 
\newcommand{\Qmdrpoolp}{\Q^{\text{2s\&d}}}
\newcommand{\Qmdrpoolintgenp}{\Q^{\text{2s\&d-ig}}}
\newcommand{\Qndrp}{\Q^{\text{1s\&d}}}
\newcommand{\Qndrintgenp}{\Q^{\text{1s\&d-ig}}}

\newcommand{\CMH}{T_{\text{CMH}}} 
\newcommand{\CMHdrpoolp}{\CMH^{\text{2s\&d}}}
\newcommand{\CMHdrpoolintgenp}{\CMH^{\text{2s\&d-ig}}}
\newcommand{\CMHdrp}{\CMH^{\text{1s\&d}}}
\newcommand{\CMHdrintgenp}{\CMH^{\text{1s\&d-ig}}}

\newcommand{\Sighat}{\hat{s}}

\newcommand{\sighatdriftsq}{\hat{\sigma}_{\text{drift}}^2}
\newcommand{\sighatdriftsqk}{\hat{\sigma}_{\text{drift-k}}^2}

\newcommand{\pbartwohat}{\hat{p}_2}
\newcommand{\pbartwohatk}{\hat{p}_{2k}}

\newcommand{\fhat}{\hat{f}}

\newcommand{\xhat}{\hat{x}}
\newcommand{\bigxhat}{\hat{X}}

\newcommand{\Ttilden}{T_n}
\newcommand{\Ttildenk}{T_{nk}}

\endlocaldefs

\begin{document}

\begin{frontmatter}

\title{
Modifying the Chi-square and the CMH test for population genetic inference: adapting to over-dispersion}
\runtitle{Chi-square and CMH test under overdispersion}

\begin{aug}
\author{\fnms{Kerstin} \snm{Spitzer}\thanksref{t1,m1,m2}\ead[label=e1]{Kerstin.Gaertner@vetmeduni.ac.at}},
\author{\fnms{Marta} \snm{Pelizzola}\thanksref{m1,m2}\ead[label=e2]{Marta.Pelizzola@vetmeduni.ac.at}}
\and
\author{\fnms{Andreas} \snm{Futschik}\corref{} \thanksref{m2,m3}
\ead[label=e3]{andreas.futschik@jku.at}}

\thankstext{t1}{previously: Kerstin Gärtner}

\affiliation{Vetmeduni Vienna\thanksmark{m1}, Vienna Graduate School of Population Genetics\thanksmark{m2} and Johannes Kepler University Linz\thanksmark{m3}}



\end{aug}

\begin{abstract}

Evolve and resequence studies provide a popular approach to simulate evolution in the lab and explore its genetic basis. 
In this context, the
chi-square test, Fishers exact test, as well as the Cochran-Mantel-Haenszel test are commonly used to infer genomic positions affected by selection from temporal changes in allele frequency. 
However, the null model associated with these tests does not match the null hypothesis of actual interest.
Indeed due to genetic drift and possibly other additional noise components such as pool sequencing, 
the null variance in the data 
can be substantially larger than accounted for by these common test statistics.
This leads to p-values that are systematically too small and therefore a huge number of false positive results. 
Even, if  the ranking rather than the actual p-values is of interest,
a naive application of the mentioned tests will give misleading results, as the 
amount of over-dispersion varies from locus to locus.
We therefore propose adjusted statistics that 
take the over-dispersion into account while keeping the formulas simple. 
This is particularly useful in genome-wide applications, where millions of SNPs can be handled with little 
computational effort.
We then apply the adapted test statistics to real data from \textit{Drosophila}, and
investigate how information from intermediate generations can be included when available.
The obtained formulas may also be useful in other situations, 
provided that the null variance  either is known or can be estimated. 
\end{abstract}


\begin{keyword}
\kwd{chi-square test}
\kwd{CMH test}
\kwd{over-dispersion}
\kwd{experimental evolution}
\kwd{evolve and resequence}
\kwd{genetic drift}
\kwd{pool sequencing}
\end{keyword}

\end{frontmatter}

\section{Introduction}  \label{sec_Introduction}


An important question in the field of population genetics is how populations adapt to changes in their environment. Experimental evolution allows to study the adaptation under controlled conditions. If these experiments are combined with high-throughput sequencing, they are called evolve and resequencing (E\&R) experiments \citep{turner_population-based_2011}. Such experiments are carried out both on microbes and on higher organisms. Due to large 
population sizes and short generation times, microbes permit to study evolutionary processes based on newly arriving mutations. With higher organisms 
and sexual reproduction on the other hand, evolution based on standing genetic variation is usually  explored.
A major goal is to  identify genomic positions (a.k.a.\ loci) that are responsible for the adaptation. 
For this purpose, the organisms are kept over $t+1$ generations $G_0, G_1,\ldots, G_t$ under conditions that require adaptation.
Allele frequencies are obtained by sequencing the genomes at $G_0$ and $G_t$ and possibly also at some intermediate time points.
Depending on the (time and financial) budget, the organisms are sequenced individually, or as a pool in order to obtain allele frequencies for typically millions of 
single-nucleotide polymorphisms(SNPs).
Individual sequencing can also be implemented using barcoding, where barcode tags that identify the organism are added before sequencing.

Allele frequency changes over time are then tested for signals of selection.  
For this purpose, usually only bi-allelic SNPs are considered. Indeed, multi-allelic sites are rare in population data and likely caused by sequencing errors 
\citep{burke_genome-wide_2010}. Consequently frequencies of the two alleles are used 
in the base and the evolved population for each tested SNP.
 Pearson's chi-square test (for simplicity subsequently called chi-square test) is a
 very popular method for this purpose  \citep{griffin_genomic_2017}.
 Serving the same purpose, Fishers exact test is sometimes used as an alternative \citep{burke2010genome}.
 Being a generalization of the chi-square test for stratified data, the CMH test  is also often used when allele frequency data from several replicate populations are available. 
 See e.g.\,\citet{orozco-terwengel_adaptation_2012, tobler_massive_2014, nouhaud_ancestral_2016, barghi_drosophila_2017} for applications.

\citet{kofler_guide_2014} compare several methods for detecting selection by contrasting allele frequencies in different generations. Apart 
from the CMH test, they consider the pairwise summary statistic ``diffStat'' \citep{turner_population-based_2011}, an association statistic by \citet{turner_investigating_2012}, and $F_{ST}$ \citep{remolina_genomic_2012}. A comparison of receiver operator curves (ROC) for these tests shows that the CMH test performs best, i.e. has more power than the other tests considered to identify selected SNPs. 
The aforementioned methods consider two time points, i.e.\,two generations, only. However, in E\&R experiments sometimes  organisms are sequenced
also at intermediate generations, resulting in repeated measurement data. 
Further methods are available for detecting selection in this context. 
 The method of \citet{bollback_estimation_2008}, for instance, is based on a hidden Markov model (HMM). Generalizations of this approach are provided
by \citet{malaspinas_estimating_2012} and \citet{steinrucken_novel_2014}. Also, \citet{mathieson_estimating_2013} adapt HMMs to structured populations. 
The method CLEAR by \citet{iranmehr_clear_2017} uses Markov chains in a discrete state model and computes the exact likelihood for small populations. 
Ignoring spatial dependence, linked loci are modeled using composite likelihood statistics. 
A frequency increment test (FIT) based on an approximation of the allele frequency dynamics by a Gaussian process is proposed by \citet{feder_identifying_2014}. Considering all loci separately, 
\citet{topa_gaussian_2015}  also model the allele frequency trajectories by Gaussian processes, whereas
\citet{terhorst_multi-locus_2015} approximate the joint likelihood for multiple loci.
The approach by \citet{taus_quantifying_2017} is based on linear least square (LLS) regression to fit the allele frequency data to a selection model.
\citet{schraiber_bayesian_2016} estimate parameters in a Bayesian framework with Markov-chain Monte-Carlo sampling. 
Another Bayesian approach has been proposed by \citet{levy_quantitative_2015} for estimating parameters in barcoded lineages. 
Finally, the Wright-Fisher ABC method proposed by \citet{foll_wfabc_2015} applies approximate Bayesian computation (ABC).

Besides detecting loci under selection, several of the discussed approaches additionally estimate selection coefficients, often jointly 
with other parameters like the effective population size \citep[e.g.][]{bollback_estimation_2008}, 
age of alleles \citep[e.g.][]{malaspinas_estimating_2012,schraiber_bayesian_2016}, or allelic dominance \citep[e.g.][]{taus_quantifying_2017}. 
However, such methods are computationally much more demanding than simple methods like the chi-square and the CMH tests. 
Hence, the latter are still widely used when testing for selection and e.g. implemented in the software tool 
PoPoolation2 \citep{kofler_popoolation2_2011}.

When comparing the allele frequencies between pairs of samples, the null model for the classical chi-square,
Fishers exact test, and also the CMH test assumes that the probability of sampling a given allele is the same within each given pair.
However, the sampling variation is not the only component of variance relevant in E\&R experiments.
Allele frequency changes between generations happen because of genetic drift, i.e.\,due to chance. This 
increases the variance in the data noticeably unless population sizes are large enough to safely ignore drift. 
Such a situation usually occurs only when working with micro-organisms  
\citep[e.g.][]{illingworth_quantifying_2012}.
Another potential source of random variation is pool sequencing
where the obtained reads can be regarded as a sample from the DNA pool.

When the chi-square or the CMH tests are applied to data, which contain more variance than assumed by the tests (over-dispersion), the resulting values of the test statistic are too large and hence the p-values too small. 
As a consequence, selection is often inferred for loci where it is not present. In the simulations with drift and pool sequencing, that we carried out, the null hypothesis of neutrality is rejected in up to 80 \% of the cases, despite being true. Hence, the additional variance introduced by drift and pool sequencing is by no means negligible. A common procedure to account for it is calculating an empirical false discovery rate (FDR) for false positives due to drift and pool sequencing via computer simulations \citep{orozco-terwengel_adaptation_2012}. 
\citet{griffin_genomic_2017} chose another approach by applying three different statistical tests and considering only SNPs which are significant with respect to all three methods as candidate loci.

This issue is also well known in the unrelated context of complex surveys, where different strategies have been proposed to obtain 
appropriate tests of homogeneity. (See e.g.\ chapter 10 in \citet{lohr_sampling_2009}.)

Our aim is to adapt the chi-square test and the CMH test in a way that additional sources of variance are directly included in the test statistics, making computer simulations no longer necessary. Thus, we propose a method that is faster than the commonly used ones. When sequence data for intermediate generations is available, the additional information can be included into our test statistics without a considerable increase in computation time. Also in terms of power, our method performs better than other approaches. In particular, our method has considerably more power to detect selected SNPs than the classical CMH test with the empirical FDR correction \citep{orozco-terwengel_adaptation_2012}. Compared to the approach for detecting selection in \citet{taus_quantifying_2017}, which the authors describe faster than the CLEAR method by \citet{iranmehr_clear_2017} and the Wright-Fisher ABC by \citet{foll_wfabc_2015}, our method has also slightly more power, and is $10^5$ orders of magnitude faster. Hence, our method outperforms other common approaches to detect selected loci in speed and power.

In this article, we first present variants of the chi-square and the CMH tests for general underlying variances. The statistics derived in this step can be useful in all situations, where over-dispersion is present.  \\
Further, we provide specific formulas for the test statistics under scenarios with drift and pool sequencing, which are common in E\&R experiments, and seen also in other situations. In genome-wide association studies (GWAS) e.g.\,the CMH test is often used for the inference of an association between a trait and an allele variant, and when the data arise from pool sequencing \citep[e.g.][]{bastide_genome-wide_2013, endler_reconciling_2016}, our adapted test could be a good alternative.

The remainder of this article is structured as follows: The test statistics for general underlying variances are presented in section \ref{sec_AdaptedTests} and the scenarios with drift and pool sequencing are considered in section \ref{subsec_AdaptationDriftAndPool}. In section 
\ref{sec_Results} we apply the adapted tests to simulated data and examine their performance. We apply the tests to real data and present the results in section \ref{subsec_RealData}. A discussion and an outlook in section \ref{sec_Discussion} conclude this article.

\section{Adapted Tests}  \label{sec_AdaptedTests}

In this section, we generalize the chi-square and the CMH tests to work under over-dispersion
and derive explicit formulas for scenarios with drift and pool sequencing. 
Given the application in mind, our focus is on the null model of homogeneity, although our 
derivations also apply to the test of independence.

\renewcommand{\arraystretch}{1.3}
\begin{table}[htbp]
\caption{Standard contingency table used with chi-square test. Subsequent interpretation in our population genetic application: 
Entries are allele frequencies for a bi-allelic SNP taken either from two populations or from one population at two time points.
$n$ is the total number of sequencing reads (coverage), $x_{ij}$ are the reads for allele $j$ in population $i$, $x_{i+}$ is the total number of reads in \mbox{population $i$}, $x_{+j}$ is the total number of allele $j$ in both populations, \mbox{$i,j \in \{1,2\}$}. The frequencies are obtained either 
by individual  sequencing of a sample or by pool sequencing applied to the entire population.}
\centering
	\begin{tabu}{|l|[2pt]c|c|[2pt]c|}
	\tabucline{-}
	 & \,\, allele 1 \,\, & \,\, allele 2 \,\, & \\
	\tabucline[2pt]{-}
	\,\,population 1\, \, & $x_{11}$ & $x_{12}$ & \,\, $x_{1+}$ \,\,\\
	\tabucline{-}
	\,\,population 2\, \, & $x_{21}$ & $x_{22}$ & \,\, $x_{2+}$ \,\, \\
	\tabucline[2pt]{-}
	 & $x_{+1}$ & $x_{+2}$ & $n$ \\
	\tabucline{-}
	\end{tabu}
\label{TabGen22Drift}
\end{table}
\renewcommand{\arraystretch}{1}

We summarize our data in a \mbox{2x2} contingency table. Using the notation from table \ref{TabGen22Drift},
the chi-square test statistic in its standard form
is defined as
\bel{Qn_Vierfelder}
\Q := \sum_{i=1}^2 \sum_{j=1}^2 \frac{\big(x_{ij} - \frac{x_{i+}\,x_{+j}}{n}\big)^2}{\frac{x_{i+}\,x_{+j}}{n}}= \frac{n (x_{11} x_{22} - x_{12} x_{21})^2}{x_{1+}\;x_{2+}\;x_{+1}\;x_{+2}}.
\ee

As shown in appendix \ref{S_subsec_Qa} (equation \eqref{QnisQ}), we may rewrite $\Q$ as follows
\bel{QnAdapt}
\Q^a (\Sighat_1^2, \Sighat_2^2) : =\frac{(x_{11} x_{22} - x_{12} x_{21})^2}{x_{2+}^2 \Sighat_1^2 + x_{1+}^2 \Sighat_2^2} 
=\frac{\big(x_{11} - \frac{x_{1+}\,x_{+1}}{n}\big)^2}{\big(\frac{x_{2+}}{n}\big)^2 \Sighat_1^2 +  \big(\frac{x_{1+}}{n}\big)^2 \Sighat_2^2}, 
\ee
with $\Sighat_1^2 := x_{1+}\frac{x_{+1}}{n}\frac{x_{+2}}{n}$, and $\Sighat_2^2:= x_{2+}\frac{x_{+1}}{n}\frac{x_{+2}}{n}$.

In order to adapt the test to models that involve different variances, we may now replace
$\Sighat_1^2$ and $\Sighat_2^2$ by consistent estimators of $\Var(x_{11})$ and $\Var(x_{21})$. 
As for the classical chi-square test (see appendix \ref{S_subsec_Qa}, in particular equation \eqref{ContMap2TriIn}), $\Q^a (\Sighat_1^2, \Sighat_2^2)$ 
converges in distribution to a $\chi^2$-distribution with 1 degree of freedom under the null hypothesis of homogeneity.

The CMH test is based on a 2x2x$\kstar$ contingency table, where the $\kstar$ partial 2x2 tables are assumed to be independent. 
We use the same notation as for the chi-square test and indicate a variable belonging to the $k^{th}$ partial table with the additional index $k$.
The null hypothesis is that both true proportions within each partial table are the same, i.e.\,the odds ratio in each partial table equals 1 \citep{mcdonald_handbook_2014}. 
The classical CMH test  statistic is 
\bel{CMHStatA}
\CMH := \frac{\big(\sum_{k=1}^\kstar (x_{11k} - \frac{x_{1+k}\,x_{+1k}}{n_k})\big)^2}{\sum_{k=1}^\kstar \frac{x_{1+k}\,x_{+1k}\,x_{2+k}\,x_{+2k}}{n_k^2 (n_k -1)}},
\ee 
see \citet{agresti_categorical_2002} \footnote{\label{MHOpC} This is the test statistic as proposed by Mantel and Haenszel, which is commonly considered for the CMH test. The statistic proposed by Cochran differs by the factor $\frac{1}{n_k}$ instead of $\frac{1}{n_k-1}$ in each term of the denominator. Asymptotically this difference is negligible. }.

Analogous to the chi-square test, one can adapt the CMH test to general underlying variances. 
As a first step, we write the test statistic of the CMH test as
\bel{CMHap}
\CMH^a\big(\Sighat_{1k}^2, \Sighat_{2k}^2; k=1,\ldots,\kstar\big)
:= \frac{\big(\sum_{k=1}^\kstar (x_{11k} - \frac{x_{1+k}\,x_{+1k}}{n_k})\big)^2}{\sum_{k=1}^\kstar \big((\frac{x_{2+k}}{n_k})^2 \Sighat_{1k}^2 + (\frac{x_{1+k}}{n_k})^2 \Sighat_{2k}^2\big)}.
\ee
and insert $x_{1+k}\frac{x_{+1k}}{n_k}\frac{x_{+2k}}{n_k-1}$ for $\Sighat_{1k}^2$ and $x_{2+k}\frac{x_{+1k}}{n_k}\frac{x_{+2k}}{n_k-1}$ for $\Sighat_{2k}^2$, $k=1,\ldots,\kstar$. (See appendix \ref{S_subsec_CMHa}.)
As with the chi-square test, the formula assumes one sampling step only, which is not appropriate for more complex models.
Again, however,  $\Sighat_{1k}^2$ and $\Sighat_{2k}^2$ can be replaced by 
consistent estimators of $\Var(x_{11k})$ and $\Var(x_{21k})$, $k=1,\ldots,\kstar$.  
In the next section we present suitable variance estimators for situations with drift and pool sequencing.

\subsection{Adaptation of the tests to drift and pool sequencing} \label{subsec_AdaptationDriftAndPool}

\begin{figure}[H]
\subfigure[one sampling step
  \label{SamplingSchemeOneSampling}]{\includegraphics[width=0.95\textwidth]{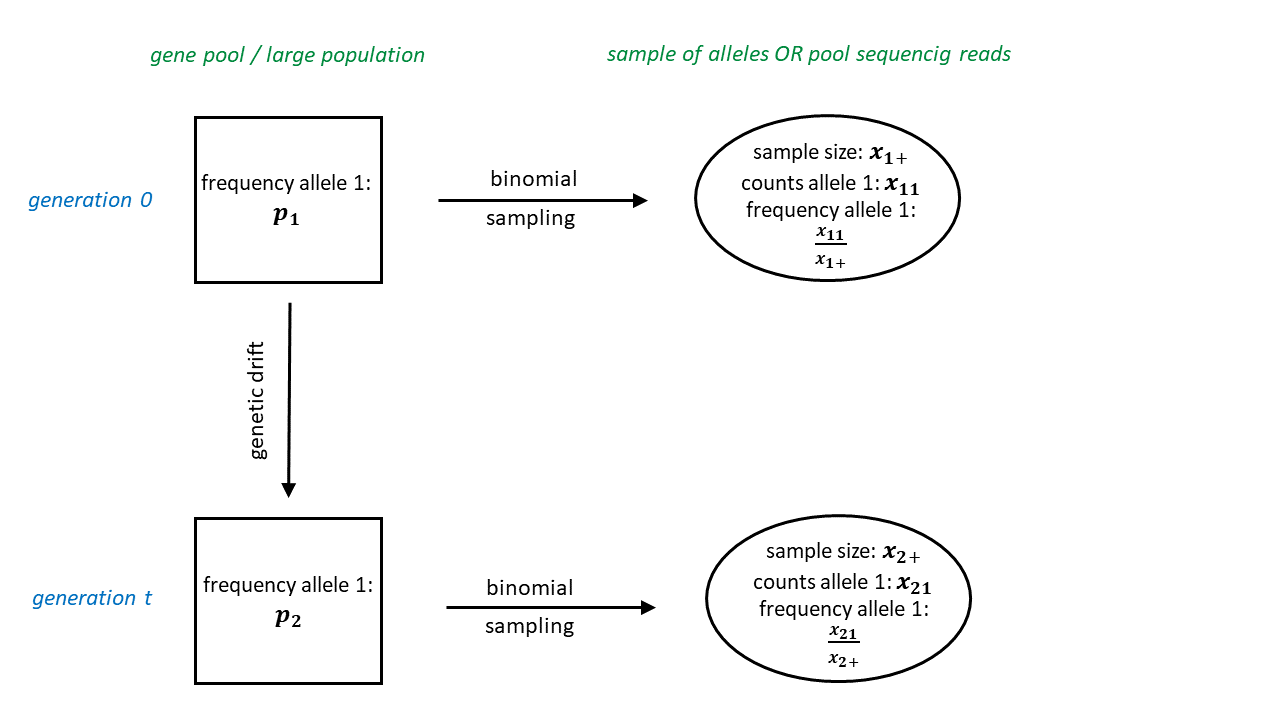}}\hfill
\subfigure[two sampling steps
  \label{SamplingSchemeTwoSampling}]{\includegraphics[width=0.95\textwidth]{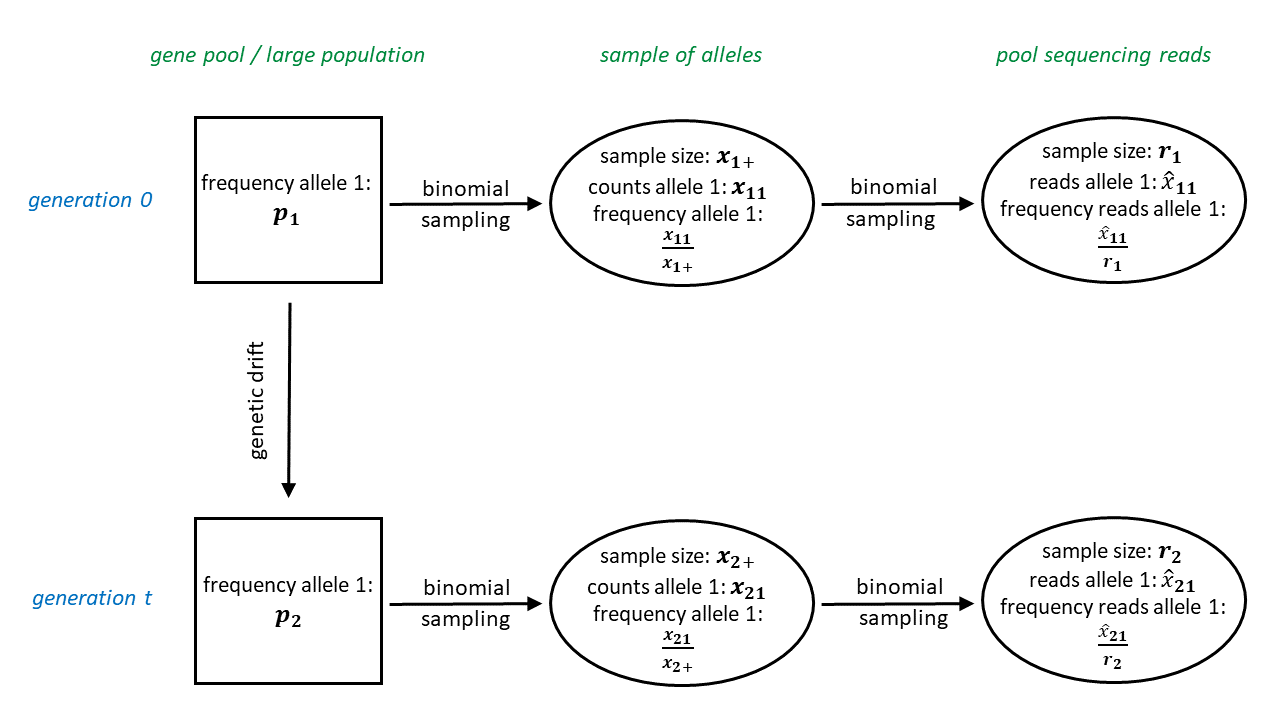}} 
\caption{Sampling schemes for the scenarios with genetic drift and one sampling step, shown in (a), or genetic drift and two sampling steps, shown in (b).}
\label{SamplingScheme}
\end{figure}

We first focus on the chi-square test and assume that allele frequency data is available for a single population at two time points.
Under the null hypothesis, the population allele frequency at the later time point $p_2$  for a given bi-allelic SNP arises from $p_1$ according to the 
Wright-Fisher model of genetic drift \citep[see e.g.\ chap.\ 3 in][]{ewens_mathematical_2004}. Therefore 
$p_2$ is modeled as a random variable. Usually one cannot observe $p_1$ and $p_2$ directly.  
Indeed both quantities are frequently estimated from population samples. If experimenters use pool sequencing as a further sampling step, we model this by binomial sampling 
as e.g.\,in \citep{waples_generalized_1989} or \citep{jonas_estimating_2016}. 
The random sample size is known as coverage, and the success probability is taken as the frequency of allele 1 in the underlying DNA material. 
If only a sample of the population is sequenced, we assume again binomial sampling for simplicity: An extension to hypergeometric sampling is straightforward.

Typically genomic selection is taken as alternative hypothesis that leads to differences between our estimates for $p_1$ and $p_2$ that cannot be explained by sampling and drift.

Figure \ref{SamplingSchemeOneSampling} summarizes the scenario with drift and one underlying sampling step, which is either taking a sample from the population for (individual) sequencing, or applying pool sequencing (to the whole population). 
The scenario with drift and two sampling steps (sampling from the population and pool sequencing) is outlined in figure \ref{SamplingSchemeTwoSampling}.

Table \ref{TabGen22Drift} summarizes the notation for a contingency table based on one sampling step (sampling from the population or pool sequencing), while table \ref{TabGen22Pool} is for two sampling steps (sampling plus pool sequencing). Note that in the scenario with two sampling steps, 
only the sample sizes $x_{1+}$ and $x_{2+}$ but not the allele frequencies are known for the first step. 
In the situations where one population descends from another, population 1 is the base population and population 2 is the evolved population.

\renewcommand{\arraystretch}{1.3}
\begin{table}[h]
\caption{Allele frequencies for a bi-allelic SNP taken either from two populations or from one population at two time points, assuming two underlying sampling steps. $m$ is the total number of sequencing reads (coverage) for the considered SNP, $\xhat_{ij}$ are the reads for allele $j$ in population $i$, $r_i$ is the total number of reads in \mbox{population $i$}, $\xhat_{+j}$ is the total number of allele $j$ in both populations, \mbox{$i,j \in \{1,2\}$}. }
\centering
	\begin{tabu}{|l|[2pt]c|c|[2pt]c|}
	\tabucline{-}
	 & \,\, allele 1 \,\, & \,\, allele 2 \,\, & \\
	\tabucline[2pt]{-}
	\,\,population 1\,\, & $\xhat_{11}$ & $\xhat_{12}$ &\,\, $r_1$ \, \\
	\tabucline{-}
	\,\,population 2 \,\,& $\xhat_{21}$ & $\xhat_{22}$ & \,\, $r_2$ \,\\
	\tabucline[2pt]{-}
	 & $\xhat_{+1}$ & $\xhat_{+2}$ & $m$ \\
	\tabucline{-}
	\end{tabu}
\label{TabGen22Pool}
\end{table}
\renewcommand{\arraystretch}{1}

Conditional on $p_1$, the variance in allele frequency due to drift after $t$ generations can be calculated as 
\bel{VarDrift}
p_1\,(1-p_1)\,\Big(1 - \big(1 - \frac{1}{2\,N_e} \big)^t \Big),
\ee
where $p_1$ is the frequency of allele 1 in the base population and $2 \,N_e$ is the effective population size considering gametes 
\citep{falconer_introduction_1960}.

If we have data from $\kstar$ replicate populations, the above considerations 
hold analogously for every replicate $k \in \{1,\ldots,\kstar\}$ and we indicate the $k^{\text{th}}$ replicate by an additional index $k$ such as in $p_{ik}$, $i\in\{1,2\}$.

We derive variance estimators for the described scenarios in Appendix \ref{S_subsec_ExplicitVars}. 
In table \ref{OurVarianceEstimatorsQm}, we present our estimators for $\Var(x_{11})$ and $\Var(x_{21})$, and in table \ref{OurVarianceEstimatorsCMH} 
for $\Var(x_{11k})$ and $\Var(x_{21k})$. 
To obtain the adapted test statistics for the different scenarios, they can 
be inserted for $\Sighat_1^2$ and $\Sighat_2^2$ in \eqref{QnAdapt}, or $\Sighat_{1k}^2$ and $\Sighat_{2k}^2$ in \eqref{CMHap} respectively.
The proposed formulas use estimators $\pbartwohat$ and $\pbartwohatk$ of $\E[p_2|p_1]$ and $\E[p_{2k}|p_{1k}]$. Also,  $\sighatdriftsq$  and $\sighatdriftsqk$ are
estimators of $\Var(p_2|p_{1})$ and $\Var(p_{2k}|p_{1k})$. Choices for these quantities are discussed below.

\small
\renewcommand{\arraystretch}{2}
\begin{table}[htbp]
\caption{Estimators $\Sighat_{1}^2$ and $\Sighat_{2}^2$ of $\Var(x_{11})$ and $\Var(x_{21})$ for different scenarios.}
\centering
\begin{tabu}{|l|[2pt]c|c|}
\tabucline{-}
& \boldmath $\Sighat_{1}^2$ & \boldmath $\Sighat_{2}^2$  \\
\tabucline[2pt]{-}
1 sampling step$^\star$  & $x_{1+}\frac{x_{+1}}{n}\frac{x_{+2}}{n}$  &  $x_{2+}\frac{x_{+1}}{n}\frac{x_{+2}}{n}$ \\
\tabucline{-}
1 sampling step, drift & $\frac{x_{11}\,x_{12}}{x_{1+}}$ & $x_{2+}\big(\pbartwohat \big(1 - \pbartwohat) + \big(x_{2+} - 1\big)	\sighatdriftsq\big)$ \\
\tabucline{-}
2 sampling steps   & $\frac{\xhat_{11}\,\xhat_{12}}{r_1}\big(1 + \frac{r_1-1}{x_{1+}}\big)$ & $\frac{\xhat_{21}\,\xhat_{22}}{r_2}\big(1 + \frac{r_2-1}{x_{2+}}\big)$\\
\tabucline{-}
2 sampling steps, drift  & $\frac{\xhat_{11}\,\xhat_{12}}{r_1}\big(1 + \frac{r_1-1}{x_{1+}}\big)$ & $r_2\big(\pbartwohat  (1-\pbartwohat) \big(1 + \frac{r_2-1}{x_{2+}}\big) + (r_2-1)\,\frac{x_{2+}-1}{x_{2+}} \sighatdriftsq \big)$\\ 
\tabucline{-}
\multicolumn{3}{l}{$^\star$ This is the situation of the classical chi-square test.}
\end{tabu}
\label{OurVarianceEstimatorsQm}
\end{table}
\renewcommand{\arraystretch}{1}
\normalsize

\small
\renewcommand{\arraystretch}{2}
\begin{table}[htbp]
\caption{Estimators $\Sighat_{1k}^2$ and $\Sighat_{2k}^2$ of $\Var(x_{11k})$ and $\Var(x_{21k})$ for different scenarios.}
\centering
\begin{tabu}{|l|[2pt]c|c|}
\tabucline{-}
& \boldmath $\Sighat_{1k}^2$ & \boldmath $\Sighat_{2k}^2$  \\
\tabucline[2pt]{-}
1 sampling step$^\star$ & $ x_{1+k}\frac{x_{+1k}}{n_k}\frac{x_{+2k}}{n_k-1}$  &  $x_{2+k}\frac{x_{+1k}}{n_k}\frac{x_{+2k}}{n_k-1}$ \\
\tabucline{-}
1 sampling step, drift  & $\frac{x_{11k}\,x_{12k}}{x_{1+k}}$ & $x_{2+k}\big(\pbartwohatk\big(1 - \pbartwohatk) + \big(x_{2+k} - 1\big)\sighatdriftsqk\big)$ \\
\tabucline{-}
2 sampling steps   & $\frac{\xhat_{11k}\,\xhat_{12k}}{r_{1k}}\big(1 + \frac{r_{1k}-1}{x_{1+k}}\big)$ & $\frac{\xhat_{21k}\,\xhat_{22k}}{r_{2k}}\big(1 + \frac{r_{2k}-1}{x_{2+k}}\big)$\\
\tabucline{-}
2 sampling steps, drift & $\frac{\xhat_{11k}\,\xhat_{12k}}{r_{1k}}\big(1 + \frac{r_{1k}-1}{x_{1+k}}\big)$ & $r_{2k}\big(\pbartwohatk  (1-\pbartwohatk) \big(1 + \frac{r_{2k}-1}{x_{2+k}}\big) + (r_{2k}-1)\,\frac{x_{2+k}-1}{x_{2+k}} \sighatdriftsqk \big)$\\ 
\tabucline{-}
\multicolumn{3}{l}{$^\star$ This is the situation of the classical CMH test.}
\end{tabu}
\label{OurVarianceEstimatorsCMH}
\end{table}
\renewcommand{\arraystretch}{1}
\normalsize

Notice that different models may apply at different time points. If an experiment involves for instance individual sequencing of a sample from the base population and pool sequencing of a sample of the evolved population, the variance estimators should be chosen accordingly:
One would take \eqref{QnAdapt} as test statistic with $\Sighat_1^2$ replaced by $\frac{x_{11}\,x_{12}}{x_{1+}}$
and $\Sighat_2^2$ replaced by $r_2\big(\pbartwohat  (1-\pbartwohat) \big(1 + \frac{r_2-1}{x_{2+}}\big) + (r_2-1)\,\frac{x_{2+}-1}{x_{2+}} \sighatdriftsq \big)$.

A simple estimator for $\E[p_2|p_1]$ is $\frac{x_{11}}{x_{1+}}$ or $\frac{\xhat_{11}}{r_{1}}$, depending on the number of underlying sampling steps. However, our simulations show that often the distribution of the corresponding p-values is closer to a uniform distribution \mbox{on $[0,1]$} when $\E[p_2|p_1]$ is estimated as 
\bel{p2h}
\frac{\frac{x_{11}}{x_{1+}} + \frac{x_{21}}{x_{2+}}}{2} \quad \text{ or } \quad \frac{\frac{\xhat_{11}}{r_{1}} + \frac{\xhat_{21}}{r_{2}}}{2} \text{ respectively}.
\ee 

For a consistent variance  estimator of  
$p_2$ after  $t$ generations of drift, $\sighatdriftsq$, we can approximate \eqref{VarDrift} by
\bel{driftest}
\frac{x_{11}\,x_{12}}{x_{1+}^2}\,\Big(1 - \big(1 - \frac{1}{2\,N_e} \big)^t \Big) \quad \text{ or } \quad  \frac{\xhat_{11}\,\xhat_{12}}{r_{1}^2}\,\Big(1 - \big(1 - \frac{1}{2\,N_e} \big)^t \Big) \text{ respectively}.
\ee

When sequence data for intermediate generations between 0 and $t$ are available, we can use this additional information for the estimation of $\E[p_2|p_{1}]$ and $\Var(p_2|p_{1})$. \\
Let $t_1=0, t_2, \ldots, t_\gamma\,=\,t$ be the generations for which sequence data is available, and let $p_1=f_1, f_2, \ldots, f_\gamma=p_2$ be the corresponding population frequencies of allele 1. 
Estimating these frequencies in each generation by the corresponding relative sample frequencies $\fhat_1, \ldots, \fhat_\gamma$,
we may proceed as in \eqref{p2h} and use this additional information to estimate $\E[p_2|p_1]$ by 
\bel{p2hig}
\frac{\sum_{i=1}^\gamma \fhat_i}{\gamma}.
\ee

Extending \eqref{driftest}, the drift variance may also be estimated  as
\bel{VarDriftintgen}
\sum_{i=1}^{\gamma-1} \fhat_i\,(1-\fhat_i)\,\Big(1 - \big(1 - \frac{1}{2\,N_e} \big)^{t_{i+1}-t_i} \Big).
\ee

Analogous estimators may be used for the estimation of $\E[p_{2k}|p_{1k}]$ and $\Var(p_{2k}|p_{1k})$ in the situation with replicate populations.

We explore the behavior of the adapted test statistics by computer simulations and present the results in the following section. Hereby we focus on the scenario with two sampling steps and genetic drift, since there the additional variance is the largest. \\

We introduce the following notation: 
If we don't have data for intermediate generations, we estimate $\E[p_2|p_{1}]$ and $\Var(p_2|p_{1})$, as well as  $\E[p_{2k}|p_{1k}]$ and $\Var(p_{2k}|p_{1k})$ by the estimators given in \eqref{p2h} and \eqref{driftest} or their analogs for the $k^{\text{th}}$ of $\kstar$ replicates. With drift and one sampling step we denote the adapted tests then  $\Qndrp$ and $\CMHdrp$, with drift and two sampling steps we name them $\Qmdrpoolp$ and $\CMHdrpoolp$. 
If data for intermediate generations is available, we apply the estimators \eqref{p2hig} and \eqref{VarDriftintgen} or their analogs for the $k^{\text{th}}$ replicate.  In the case of drift, 
we denote the adapted tests by 
$\Qndrintgenp$ resp.\ $\CMHdrintgenp$  (one sampling step), and $\Qmdrpoolintgenp$ resp.\ $\CMHdrpoolintgenp$ (two sampling steps).

\section{Simulation Results}  \label{sec_Results}

We carried out extensive simulations in \textit{R} \citep{r-core-team_r_2018} in order to explore the behavior of the adapted tests described in the previous section. We simulated genetic drift using the package \textit{poolSeq} \citep{taus_quantifying_2017}. When we encountered loci with frequency 0 for one allele in the base population but a positive frequency in a later generation, we changed the allele count from 0 to 1 in order to always obtain a well-defined test statistic. When these situations occur with real data, there are different possible explanations: Either a mutation arose in a later generation and the allele really was not present before, or the frequency of the respective allele is low, but not 0, in the base population and the allele was just by chance not sampled or amplified in the sequencing process. Since mutation rates are usually low over such a time span \citep{burke_genome-wide_2010}, the latter scenario is the more likely one. Finally, if the frequency in the later generation is very low, the nonzero frequency may also be due to a sequencing error.
Overall, our method to deal with this phenomenon seems to be a pragmatic compromise. 

  We first provide results under the null hypothesis, then we examine the power of our adapted test statistics. 
After that we compare our adapted tests to other state-of-the-art methods. 
We first simulate allele frequencies in generation 0 uniformly distributed on $[0,1]$
to give all possible true allele frequencies the same weight.
At the end of the section we also consider an allele frequency distribution in generation 0 that resembles the one encountered in our  experimental data.

\subsection*{Null distribution} \label{subsubsec_NullDistrib}
In our simulations we set $N_e=300$, and used 1000 as sample size of alleles that were sequenced at generations 0 and 60; 
the sequencing coverage was chosen Poisson distributed with mean 80. These parameter choices were motivated by the real data for \textit{Drosophila} taken from \citep{barghi_drosophila_2017} and discussed in section \ref{subsec_RealData}. 

\begin{figure}[h]
\subfigure[$\CMH$
  \label{HistUnifCmhClas}]{\includegraphics[width=0.49\textwidth]{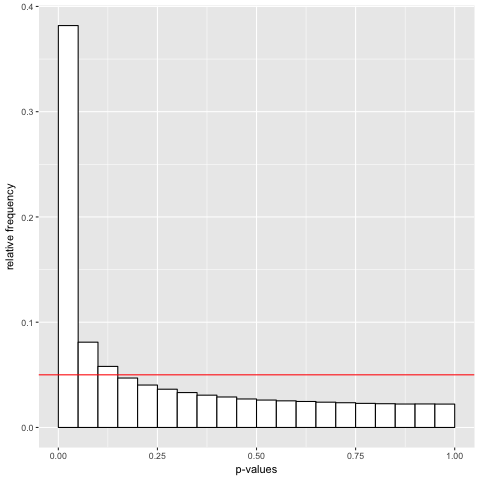}}\hfill
\subfigure[$\CMHdrpoolp$
  \label{HistUnifCmhA}]{\includegraphics[width=0.49\textwidth]{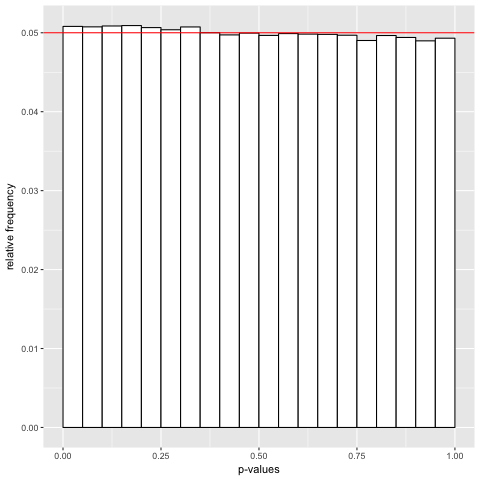}} 
\caption{Distribution of the p-values belonging to the classical CMH test $\CMH$ (a) and $\CMHdrpoolp$ (b). Red line indicates frequency of $5\%$. Simulation setup: $10^6$ neutral loci with true allele frequencies in base population uniformly distributed on $[0,1]$, $N_e=300$, allele sample size 1000, pool sequencing with coverage Poisson distributed with mean 80, sequence data for generations 0 and 60, 3 replicate populations.}
\label{HistUnifCmhClasA}
\end{figure}

To control the type I error it is desirable that the p-values of a test are uniformly distributed on $[0,1]$ or at least stochastically larger than uniform if the null hypothesis is true. Indeed, under the neutral Wright–Fisher model we observe that the distribution of the p-values belonging to the adapted tests is close to a uniform distribution and that the tests control the 5 \% significance level. In contrast, the non-adapted tests show a huge excess of small p-values. Based on $10^6$ simulated loci, figure \ref{HistUnifCmhClasA} displays the distribution of the p-values for the classical CMH test with 3 replicates and the test adapted to drift, sampling, and pool sequencing $\CMHdrpoolp$. For the adapted chi-square test $\Qmdrpoolp$, the distribution of the p-values is slightly further from a uniform distribution but controls the 5\% significance threshold (appendix figure \ref{HistUnifQmClasA}). We obtain similar results in situations with drift and only one sampling step (appendix, figure \ref{HistUnifDrift}). With two underlying sampling steps but without drift, 
a situation occurring e.g.\ in GWAS using pool sequencing data \citep[e.g.][]{bastide_genome-wide_2013, endler_reconciling_2016}, the adapted tests show again an improved performance (appendix, figure \ref{HistUnifPool}).

\begin{figure}[h]
\subfigure[$\Qmdrpoolp$
  \label{ConservativenessQmdrpool}]{\includegraphics[width=0.49\textwidth]{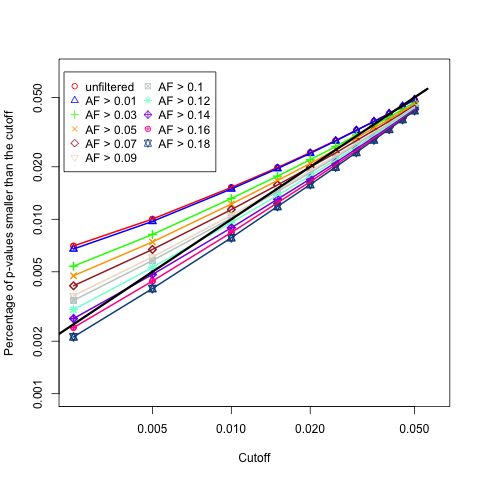}}\hfill
\subfigure[$\CMHdrpoolp$
  \label{ConservativenessCMHdrpool}]{\includegraphics[width=0.49\textwidth]{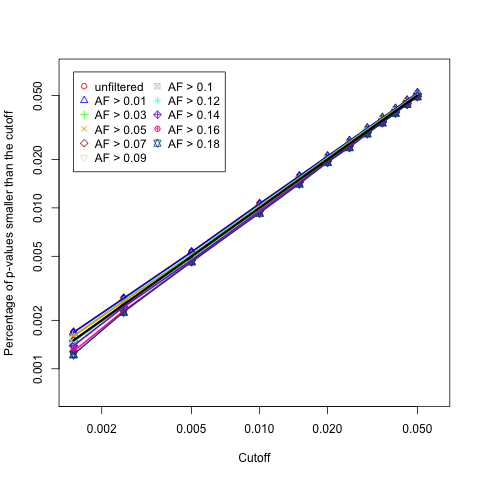}} 
\caption{Percentage of loci with p-value smaller than cut-off value against cut-off value for different minimum values of the allele frequencies of both alleles in the base population, in (a) for $\Qmdrpoolp$, and in (b) for $\CMHdrpoolp$. The black line is the angle bisector. Simulation setup: $10^6$ neutral loci with true allele frequencies in base population uniformly distributed on $[0,1]$, $N_e=300$, allele sample size 1000, pool sequencing with coverage Poisson distributed with mean 80, sequence data for generations 0 and 60, 5 replicate populations in (b).}
\label{Conservativeness}
\end{figure}

In genome-wide applications and corresponding corrections for multiple testing, also significance levels much below 0.05 become important.
To check the type I error control in such a situation,
we did a simulation analysis for $10^6$ loci based on sequence data with the parameter choices as above
 (but 5 replicate populations in case of the CMH test).
As shown in figure \ref{Conservativeness}, especially $\Qmdrpoolp$ turns out to be anti-conservative
for very small significance levels. The red lines correspond to the full, unfiltered data sets. \\ 
To understand this issue, it should be noted that the distributional approximations involved in these tests are less reliable for
loci with a very small or very high allele frequency. 

An obvious remedy would be to introduce a
threshold value $\gamma$ and only consider loci with frequency of allele 1 as well as allele 2 larger than $\gamma$ in the base population. 
According to figure \ref{Conservativeness}, filtering has the desired effect, if $\gamma$ is chosen large enough.
Indeed, the asymptotic $\chi^2$-distribution of the test statistic under the null model for the chi-square test as well as for the CMH test 
is based on the normal approximation of the binomial distribution, which is more accurate for intermediate frequencies for finite samples.

One disadvantage of this approach is that by filtering out SNPs with small and large allele frequencies, we exclude a lot of potentially selected loci.
We therefore explore also other approaches to resolve the issue:
If sequence data is available not only from two time points but also from intermediate generations, we can modify the adapted tests by taking the additional information into account, resulting in the test statistics $\Qndrintgenp$,  $\Qmdrpoolintgenp$, $\CMHdrintgenp$, and $\CMHdrpoolintgenp$, see \mbox{section \ref{subsec_AdaptationDriftAndPool}}.
Simulations with the same parameters as in figure \ref{Conservativeness}, but additionally with sequence data every 10 generations, show that $\Qmdrpoolintgenp$ and $\CMHdrpoolintgenp$ hold the 5\% level or are conservative for all significance levels without filtering out small and large allele frequencies (appendix, figure \ref{ConservativenessIntgen}).

If time series data is not available, a p-value correction may be applied by fitting the distribution $F_p$ of p-values 
simulated under the null hypothesis and transforming the p-values to uniform using $F_p(\cdot)$.
In appendix \ref{S_correc_P} we propose a parametric choice of $F_p$  for correcting small (potentially significant) p-values, and show that it 
leads to a substantial improvement of the null distribution of p-values obtained with $T^{2s\&d}_{\chi^2}.$

\subsection*{Power} \label{subsubsec_Power}

We additionally carried out simulations involving $10^5$ selected loci in order to examine the power of the adapted tests at a significance threshold of $\alpha=0.05$.
We first consider a realistic set of standard parameter choices: 
$N_e=300$, sample size: 1000, coverage: 100, and sequence data available for generations 0 and 60. 
To also investigate the influence of these model parameters, we considered a set of alternative values for each of them.

\begin{figure}[H]
\subfigure[$\Qmdrpoolp$, 60 generations
  \label{PowerDepSQm60}]{\includegraphics[width=0.49\textwidth]{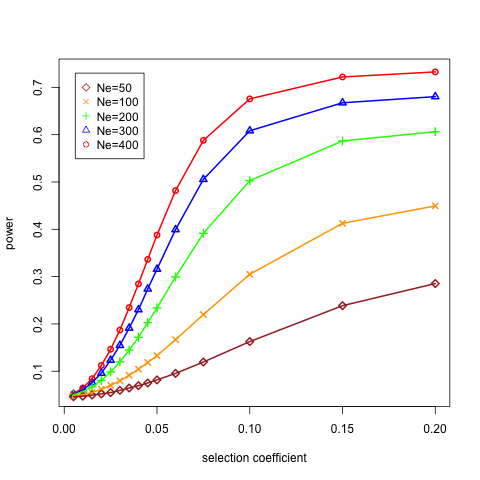}}\hfill
\subfigure[$\CMHdrpoolp$, 60 generations
  \label{PowerDepSCMH60}]{\includegraphics[width=0.49\textwidth]{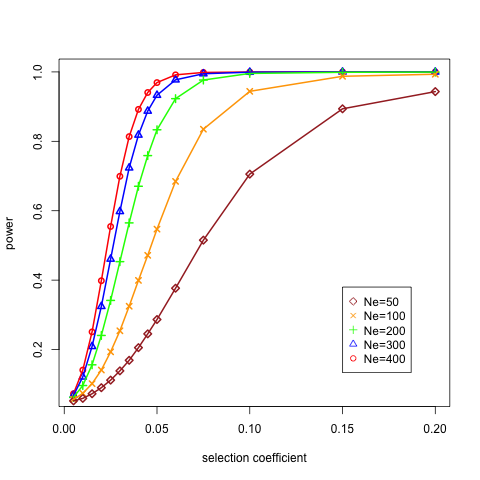}} 
\subfigure[$\Qmdrpoolp$, 20 generations
  \label{PowerDepSQm20}]{\includegraphics[width=0.49\textwidth]{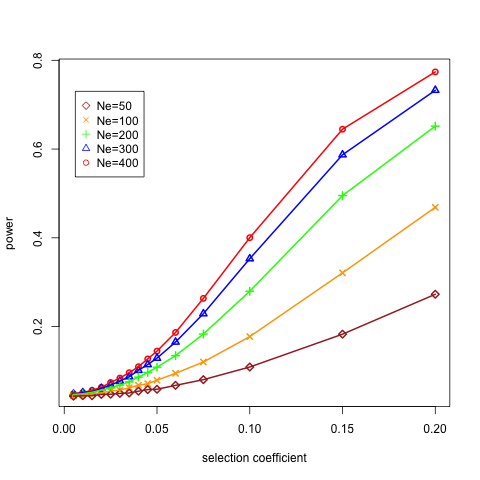}}
\subfigure[$\Qmdrpoolp$, 200 generations
  \label{PowerDepSQm200}]{\includegraphics[width=0.49\textwidth]{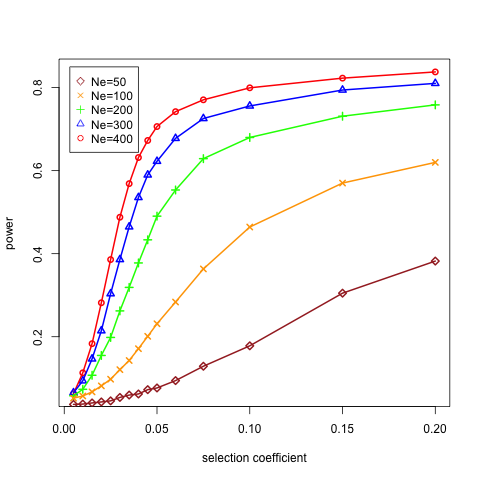}}
\caption{Power vs selection coefficients for different values of the effective population size $N_e$, in (a), (c), (d) for $\Qmdrpoolp$, and in (b) for $\CMHdrpoolp$. Simulation setup: $10^5$ loci for each selection coefficient with true allele frequencies in base population uniformly distributed on $[0,1]$, allele sample size 1000, pool sequencing with coverage 100, 5 replicate populations in (b);  (a) and (b) sequence data generations 0 and 60, (c) sequence data generations 0 and 20, (d) sequence data generations 0 and 200.}
\label{PowerDepS}
\end{figure}

Not surprisingly, the power of $\CMHdrpoolp$ is higher than the power of $\Qmdrpoolp$ because the amount of information increases with more replicates. As also expected, the power increases with the selection coefficient $s$ and the effective population size $N_e$. \mbox{Figure \ref{PowerDepSQm60}} shows the power of $\Qmdrpoolp$ for different values 

of $s$ and $N_e$. 
Figure \ref{PowerDepSCMH60} displays the same for $\CMHdrpoolp$ with 5 replicates. Comparing figures \ref{PowerDepSQm20} and \ref{PowerDepSQm200}, 
which are for the same scenario as in figure \ref{PowerDepSQm60} but with 20 and 200 generations of evolution, we see that for small selection coefficients the power increases with the number of generations. This is since more generations of evolution lead to larger frequency differences 
between base and evolved populations unless the selection coefficients are large. Under strong selection many alleles soon reach 
frequency close to 1 (fixation), and hence the signal of selection does not become stronger anymore with more generations. On the other hand, the drift variance that we calculate for the denominator of the test statistic increases with every generation, which reduces the power. 
The gain in power due to more generations of evolution and the loss in power due to fixation may cancel each other out, leading to the plateauing effect at around $0.8$ we observe in figure \ref{PowerDepSQm200}. The impact of the number of generations on the power is qualitatively the same for $\CMHdrpoolp$. (appendix, figure \ref{PowerDepS_supCMH}).

\begin{figure}[h]
\subfigure[$\Qmdrpoolp$
  \label{PowerDepSizeQm}]{\includegraphics[width=0.49\textwidth]{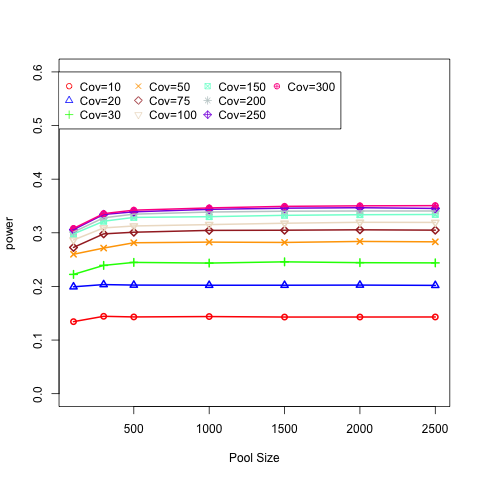}}\hfill
\subfigure[$\CMHdrpoolp$
  \label{PowerDepSizeCMH}]{\includegraphics[width=0.49\textwidth]{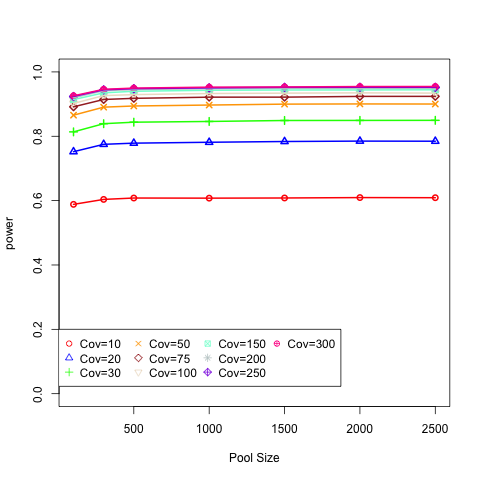}} 
\caption{Power vs sample size of alleles (pool size) for different pool sequencing coverages, in (a) for $\Qmdrpoolp$, and in (b) for $\CMHdrpoolp$. Simulation setup: $10^5$ loci for each value of the pool size with true allele frequencies in base population uniformly distributed on $[0,1]$, $N_e=300$, selection coefficient 0.05, sequence data for generations 0 and 60, 5 replicate populations in (b).}
\label{PowerDepSize}
\end{figure}

The influence of sample size and coverage is shown in figure \ref{PowerDepSize}, in (a) for $\Qmdrpoolp$, and in (b) for $\CMHdrpoolp$.  
The power increases with the sample size and with the coverage. Since the coverage values are an order of magnitude smaller than the values for the sample size, the effect is much more pronounced for the coverage.

The power of $\Qmdrpoolintgenp$ and $\CMHdrpoolintgenp$ is similar to the power of $\Qmdrpoolp$ and $\CMHdrpoolp$, see  table \ref{CompareMethodsTab}. In general, $\Qmdrpoolintgenp$ and $\CMHdrpoolintgenp$ are affected in the same way by the different parameters as $\Qmdrpoolp$ and $\CMHdrpoolp$ (not shown).

\subsection*{Method comparison} \label{subsubsec_MethodComparison}

We compared the performance of our adapted test statistics to other state-of-the-art methods. As summaries we considered the type I error, the power, and the run time. We looked at the classical chi-square test with an empirical FDR correction \citep{orozco-terwengel_adaptation_2012} and the LLS approach for detecting selection of \citet{taus_quantifying_2017}. As stated in  \citet{taus_quantifying_2017}, the LLS method is faster than the methods CLEAR by \citet{iranmehr_clear_2017} and Wright-Fisher ABC by \citet{foll_wfabc_2015}. Still the computation times with LLS method are high and we restricted our simulations therefore here to $10^4$ loci of which 10 \% are under selection. The other parameters were chosen again $N_e=300$, allele sample size 1000, coverage Poisson distributed with mean 80, 60 generations of drift.

Compared with the LLS method by \citet{taus_quantifying_2017}, $\Qmdrpoolp$ and $\Qmdrpoolintgenp$ have a similar power to detect selection while being much faster, see table \ref{CompareMethodsTab}.
The chi-square test with the empirical FDR correction \citep{orozco-terwengel_adaptation_2012} is not much slower than our tests (although the empirical FDR 
correction requires additional computations). As the test statistic does not use the correct variance terms, it performs worse, however, than our corrected tests both in terms of the power 
and the type I error. 


\renewcommand{\arraystretch}{1.3}
\begin{table}[htbp]
\caption{Comparison of type I error, power, and running times in seconds for different tests of selection. Simulation setup: $10^4$ loci with true allele frequencies in base population uniformly distributed on $[0,1]$, 10\% of the loci under selection, selection coefficients exponentially distributed with mean 0.1, $N_e=300$, allele sample size 1000, pool sequencing with coverage Poisson distributed with mean 80, sequence data for generations 0 and 60, additionally for generations 10, 20, 30, 40, 50 when $\Qmdrpoolintgenp$ and $\CMHdrpoolintgenp$ are applied, 5 replicate populations in case of $\CMHdrpoolp$ and $\CMHdrpoolintgenp$.} 
\centering
	\begin{tabu}{|l|c|c|c|}
	\tabucline{-}
	Test &Type I Error & Power & Time \\
	\tabucline[2pt]{-}
	$\Qmdrpoolp$ & 0.050 & 0.417 & 0.005\\
	\tabucline{-}
	$\Qmdrpoolintgenp$ & 0.046 & 0.490 & 0.369\\
	\tabucline{-}
	$\CMHdrpoolp$ with 5 replicates&  0.050 & 0.761 & 0.014\\
	\tabucline{-}
	$\CMHdrpoolintgenp$ with 5 replicates & 0.049 & 0.756 & 0.487\\
	\tabucline{-}
	$\Q$ with empirical FDR by \citep{orozco-terwengel_adaptation_2012}  &  0.374 & 0.226 & 0.144\\
	\tabucline{-}
	LLS$^\star$ \citep{taus_quantifying_2017} &  0.047 & 0.456 & 33568.010\\
	\tabucline{-}
	\end{tabu}
	\begin{flushleft}$^\star$ Assume diploids, dominance is set to 0.5, the method to estimate selection is set to "LLS" and p-values are simulated with N.pval set to 1000, which means that 1000 simulations are performed to estimate the p-values.
	\end{flushleft}
\label{CompareMethodsTab}
\end{table}
\renewcommand{\arraystretch}{1}

\subsection*{Alternative starting allele frequency distribution} \label{subsubsec_SimBeta}

In the previous simulations the true allele frequencies in the base populations were chosen uniformly distributed on $[0,1]$. In general, uniformly distributed allele frequencies are not common in natural populations. Therefore we also looked for a more realistic distribution of allele frequencies. As an example we consider the u-shaped beta distribution proposed in  \citep{jonas_estimating_2016} for the fruit fly \textit{Drosophila}. The u-shape depicts the observed excess of high and low allele frequencies in a folded site frequency spectrum where it is not known which allele is ancestral and which one is derived. 

\begin{figure}[h]
\subfigure[$\Qmdrpoolintgenp$
  \label{HistBetaQmIntgen}]{\includegraphics[width=0.49\textwidth]{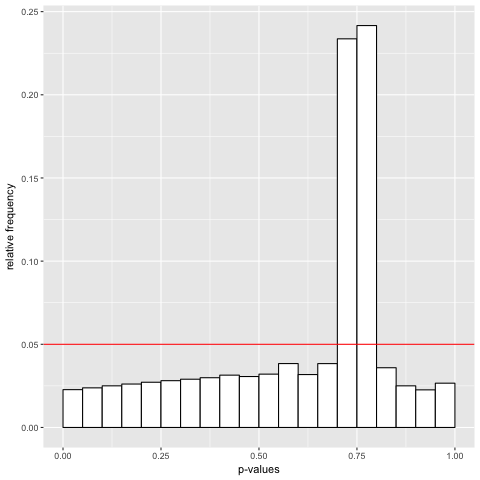}}\hfill
\subfigure[$\CMHdrpoolintgenp$
  \label{HistBetaCmhIntgen}]{\includegraphics[width=0.49\textwidth]{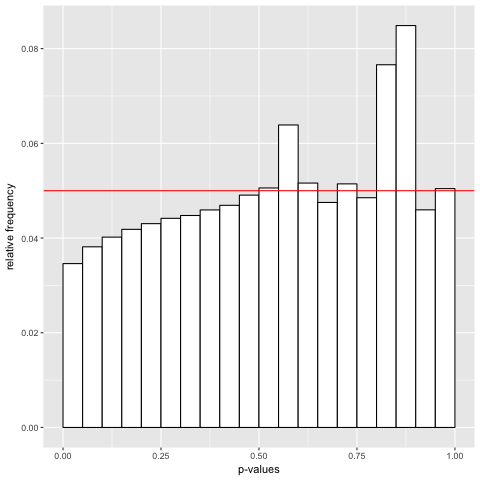}} 
\caption{Distribution of the p-values belonging to $\Qmdrpoolintgenp$ (a) and $\CMHdrpoolintgenp$ (b). Red line indicates frequency of $5\%$. Simulation setup: $10^6$ neutral loci with true allele frequencies in base population beta distributed with parameters 0.2 and 0.2, $N_e=300$, allele sample size 1000, pool sequencing with coverage Poisson distributed with mean 80, sequence data every 10 generations from generation 0 to 60, 3 replicate populations in (b).}
\label{HistBetaIntgen}
\end{figure}

We therefore chose the true allele frequencies in generation 0 from a beta-distribution with parameters 0.2 and 0.2. The other parameters remained
 unchanged compared to the previous simulations. 
As the chosen distribution produces a higher proportion of allele frequencies close to the boundaries, the distribution 
of the p-values becomes less uniform (see figure \ref{HistBetaIntgen}). The spikes are caused by discreteness phenomena occurring 
for very low starting allele frequencies.
However, $\Qmdrpoolintgenp$ and $\CMHdrpoolintgenp$ turned out to be quite conservative (even for significance levels smaller than 0.05) 
when sequence data every 10 generations is available. 
At the same time power values of the adapted tests were lower than when under uniformly distributed 
starting allele frequencies (appendix, figure \ref{BetaPower}). An explanation would be that even SNPs with positively selected 
alleles have a large probability of being lost due to drift in early generations, if the initial allele frequency is very low.
For such SNPs we do not have any power of detection.

In particular, we calculated a power of approximately 0.59 for $\CMHdrpoolp$ based on $10^5$ simulations with selection coefficient 0.1, allele sample size 1000, pool sequencing in generations 0 and 60 with coverage Poisson distributed with mean 80, and 5 replicates. When additional sequence data for generations 10, 20, 30, 40, and 50 was simulated, we obtained a power of approximately 0.60 
using $\CMHdrpoolintgenp$ as test statistic. 

\section{Application to experimental data from \textit{Drosophila}}  \label{subsec_RealData}

Here we consider data from an evolve and resequence experiment
on \textit{Drosophila simulans} as described in \citep{barghi_drosophila_2017}. 
In this publication the classical CMH test has been used to infer candidates of selection. 
Allele frequency measurements were taken from three replicate populations at generations 0 and 60. 
All flies were maintained under a cycling routine of 12 hours at 18$^\circ$C in a dark environment to mimic night and 12 hours at 28$^\circ$C with light for the day. 

In the original paper neutral simulations have been used to define a cut-off that leads to 2\% false 
positive SNPs under the simulated global null model. 
This cut-off has then been taken as a threshold for the p-values obtained with the CMH test applied to the real data, see \citep{orozco-terwengel_adaptation_2012} for a more detailed description. 
Notice, however, that with approximately 4 million SNPs \citep{barghi_drosophila_2017}, this approach will lead to approximately 80000 false positive SNPs and an unclear false discovery rate.

An advantage of our approach is that the resulting proper p-values can be combined with a standard procedure such as Bonferroni or Benjamini-Hochberg that controls for multiple testing. Another advantage of our method is that it will lead 
to a more proper ranking of the p-values, as the amount of error incurred with the classical test statistics depends on the relative magnitudes
of the variance components (drift variance, sequencing coverage and sample size) and will therefore vary between SNPs.

\begin{figure}
\includegraphics[width=.50\linewidth]{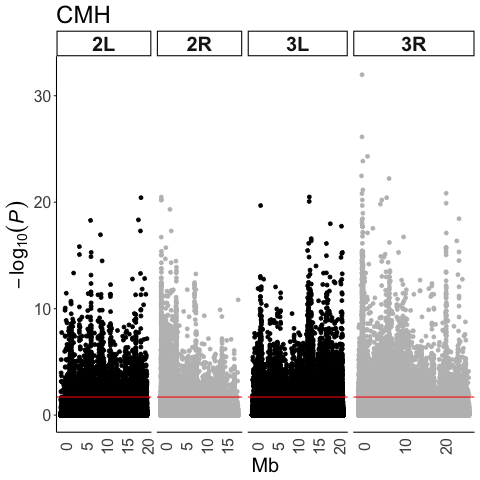}
\caption{Manhatten plot for $\CMHdrp$ applied to real data from \textit{Drosophila simulans} \mbox{from \citep{barghi_drosophila_2017}.}}
\label{fig:cmh}
\end{figure}

Since the DNA from the whole population was used in the pool sequencing step, 
we applied $\Qndrintgenp$ and $\CMHdrintgenp$ as test statistics using the model parameters specified in \citep{barghi_drosophila_2017}. 
We only present the results for the modified CMH test on the entire data set.  
A Manhattan plot
of the SNP positions versus the logarithm (base 10) of the p-values corrected for multiple testing with the Benjamini-Hochberg method is shown in figure \ref{fig:cmh}. 
The computation time needed for this analysis was only about 20 seconds on a standard laptop. We infer more significant SNPs than \citep{barghi_drosophila_2017} ($0.0049\%$ and $0.0002\%$ of the total number of tested SNPs, respectively). This result is concordant with our simulations showing that our method has higher power compared to the empirical FDR correction \citep{orozco-terwengel_adaptation_2012} which is also applied in \citep{barghi_drosophila_2017}. 
The population genetics group at the 
University of Veterinary Medicine in Vienna plans to apply our tests on
new data sets that will become available soon.

\section{Discussion}  \label{sec_Discussion}

With population genetic applications in mind, we propose modified test statistics for the chi-square and the CMH tests in scenarios with over-dispersion, 
i.e.\,more variance in the data than considered by the original tests. Compared with the classical versions of these tests that are still commonly applied, our approach does not require simulations to find a cut-off for the test statistic. Our proposed approach can also be used instead of Fishers exact test which faces the same problems in this context.
Our inclusion of proper variance terms leads to a better performance both with respect to power and type I error.
Using the classical tests does not even lead to a proper ranking of the SNPs as the amount of over-dispersion varies between the SNPs. While more sophisticated testing procedures have also been proposed, they are usually considerably more time consuming, especially when applied on a genome wide scale.

While this issue is also known in other applications (such as complex surveys), our underlying model requires a different adaptation of the test statistics. We therefore expressed the test statistics in dependence of the variances of entries in the contingency tables as a first step. 
Using this general form, suitable test statistics can be calculated in any situation with over-dispersion provided that the required variances can be properly estimated. \\
We then derived explicit formulas for the adapted test statistics for use in different types of E\&R experiments under scenarios with one or two sampling steps, resulting from actual sampling or pool sequencing, and genetic drift.

Our test statistics do not only provide a more appropriate error control, but have also a much larger power than the the classical tests combined with computer simulations such as proposed in \citet{orozco-terwengel_adaptation_2012}.

Compared with more sophisticated methods, our adapted tests have approximately the same power but need much less computation time, which is an important factor when the whole genome is scanned for traces of selection. Our tests are e.g.\,$10^5$ times faster than the LLS method by \citet{taus_quantifying_2017}. 

When sequence data is available for two time points only, our tests are not always conservative for very small allele frequencies in the base population. 
This problem is related with inaccuracies of the normal approximation of proportions close to zero and one.
In this case, we propose a correction of the p-values based on simulations under the null model. 
Notice, however, that this additional correction is usually not needed when data is available also at intermediate time points, as supported by
our simulated scenarios.

We implemented the adapted tests in an \textit{R} package called ACER, which can be downloaded on \textit{https://github.com/MartaPelizzola/ACER}. 
If many loci or even the whole genome is tested for selection, we recommend to control for the FDR by the Benjamini-Hochberg or a similar procedure.

Our results suggest that our adapted test statistics provide fast, reliable and powerful methods to detect selection. 
Hence, they have the potential to considerably facilitate the inference of selected loci in population genetics, in particular in the context of E\&R.

\section*{Acknowledgments}

We thank Neda Barghi and Christian Schlötterer for providing the data and helpful comments, as well as Thomas Taus for his input on the acceleration of the R-code. This work has been supported by the Austrian Science Fund (FWF Doctoral Program ``Vienna Graduate School of Population Genetics'', DK W1225-B20).

\bibliography{Manuscript}  

\newpage
\section*{Appendix}
\setcounter{equation}{0}
\setcounter{figure}{0}
\setcounter{table}{0}
\setcounter{page}{1}
\makeatletter
\renewcommand{\theequation}{A\arabic{equation}}
\renewcommand{\thefigure}{A\arabic{figure}}
\renewcommand{\bibnumfmt}[1]{[A#1]}
\renewcommand{\citenumfont}[1]{A#1}

\renewcommand{\thesubsection}{A\arabic{subsection}}  

\textit{Supplemental material}

\mbox{}\newline 
In our derivations we either assume $p_1$ to be not random, or do our computations conditional on $p_1$.

\subsection{Derivation of \texorpdfstring{$\Q^a (\Sighat_1^2, \Sighat_2^2)$}{}} \label{S_subsec_Qa}

We restrict our analysis here on 2x2 tables since we consider two samples\footnote{For time series data and hence more sequence samples, the samples from the intermediate time points are considered only for a variance estimator within the here described framework of two sequence samples.} and bi-allelic SNPs. However, our results could be generalized to contingency tables with larger dimensions by expressing the test statistic in dependence of the respective variances and replacing these variances by appropriate estimators, which we will do now for 2x2 tables.

We follow the rationale of chapter 14.3 in \citep{lehmann_testing_2005} for the goodness-of-fit test, but adopt it to the tests of independence and homogeneity, and derive a test statistic $\Q^a (\Sighat_1^2, \Sighat_2^2)$ for the chi-square test depending on the consistent estimators $\Sighat_1^2$ and $\Sighat_2^2$ of $\Var(X_{11})$ and $\Var(X_{21})$.
 
We assume $X_{11}$ and $X_{21}$ to be independent.
As it is usually done when testing for homogeneity, we consider the total number of alleles in each generation, $x_{1+}$ and $x_{2+}$, as known and assume 
\bel{AssBinDis}
X_{11} \sim \Binom(x_{1+}, p_1) \quad \text{ and } \quad X_{21} \sim \Binom(x_{2+}, p_2).
\ee

Note that $p_2$ will often be a random variable, such as under genetic drift. Under the null hypothesis we 
have that $E(p_2 | p_1)= p_1 =: p$. We assume that the null hypothesis is true.

We define an auxiliary variable 
\bel{DefTtilden}\Ttilden := (X_{11} - \frac{x_{1+}\,X_{+1}}{n}). \ee

With $X_{+1} = X_{11}+X_{21}$ we obtain
\bel{VarTtildenBin}
\begin{split}
\Var(\Ttilden) 
&= \Var\Big(\frac{n}{n} X_{11} - \frac{x_{1+}}{n} (X_{11}+X_{21}) \Big)  \\
&= \Var\Big(\frac{n-x_{1+}}{n} X_{11} - \frac{x_{1+}}{n} X_{21} \Big)  \\
&= \big(\frac{x_{2+}}{n}\big)^2 \Var(X_{11}) + \big(\frac{x_{1+}}{n}\big)^2 \Var(X_{21}).
\end{split}
\ee
Hence, with the consistent estimators $\Sighat_1^2$ and $\Sighat_2^2$ of $\Var(X_{11})$ and $\Var(X_{21})$, we obtain
\bel{ConvVar1}
\big(\frac{x_{2+}}{n}\big)^2 \Sighat_1^2 + \big(\frac{x_{1+}}{n}\big)^2 \Sighat_2^2 \toop \Var(\Ttilden).
\ee

Above we considered $X_{+1}$ (and hence also $X_{+2}=n-X_{+1}$) as random variable. If, however, we condition on $x_{+1}$ (and $x_{+2}$), which is e.g.\,done for the test for independence, we are in a hypergeometric framework and can estimate the variances from there. The estimator for $\Var(\Ttilden)$ from the hypergeometric framework yields a factor $\frac{1}{n-1}$ where in the binomial framework with $X_{+1}$ random there is a factor $\frac{1}{n}$. Since we are interested in the asymptotic behavior of the test statistic, this difference is negligible.

With \eqref{VarTtildenBin} and \eqref{AssBinDis} it is further
\bel{ETn0}
\begin{split}
\E[\Ttilden] 
&= \E\Big[\frac{x_{2+}}{n} X_{11} - \frac{x_{1+}}{n} X_{21} \Big] \\
&= \frac{x_{2+}}{n} \E[X_{11}] - \frac{x_{1+}}{n} \E[X_{21}] \\
&= \frac{x_{2+}}{n}\, x_{1+}\,p - \frac{x_{1+}}{n} \, x_{2+}\,p \\
&= 0
\end{split}
\ee
With the central limit theorem we can thus deduce
\bel{CLT1}
\frac{\Ttilden}{\sqrt{\Var(\Ttilden)}} \schw \Normal(0,1),
\ee
and with the continuous mapping theorem we conclude
\bel{ContMap1}
\frac{\Ttilden^2}{\Var(\Ttilden)} \schw\chi^2_1,
\ee
where $\chi^2_1$ stands for the $\chi^2$-distribution with 1 d.f. 
Finally, we obtain with \eqref{ConvVar1}, \eqref{ContMap1}, and again the continuous mapping theorem
\bel{ContMap2TriIn}
\frac{\Ttilden^2}{\big(\frac{x_{2+}}{n}\big)^2 \Sighat_1^2 + \big(\frac{x_{1+}}{n}\big)^2 \Sighat_2^2}  =\frac{\big(X_{11} - \frac{x_{1+}\,X_{+1}}{n}\big)^2}{\big(\frac{x_{2+}}{n}\big)^2 \Sighat_1^2 +  \big(\frac{x_{1+}}{n}\big)^2 \Sighat_2^2}  
=: \Q^a(\Sighat_1^2, \Sighat_2^2) \schw\chi^2_1.
\ee

Now we show that setting \mbox{$\Sighat_1^2 = x_{1+}\frac{X_{+1}}{n}\frac{X_{+2}}{n}$} and \mbox{$\Sighat_2^2 = x_{2+}\frac{X_{+1}}{n}\frac{X_{+2}}{n}$} yields the original test statistic $\Q$.\\
From \eqref{AssBinDis} we know that
\bel{VarX11AndX22}
\Var(X_{11}) = x_{1+}\,p\,(1-p) \quad \text{ and } \quad \Var(X_{21}) = x_{2+}\,p\,(1-p).
\ee
Further, $\frac{X_{+1}}{n}$ is a consistent estimator for $p$, and hence $1-\frac{X_{+1}}{n} = \frac{X_{+2}}{n}$ is a consistent estimator for $(1-p)$. This means that $x_{1+}\frac{X_{+1}}{n}\frac{X_{+2}}{n}$ and $x_{2+}\frac{X_{+1}}{n}\frac{X_{+2}}{n}$ indeed are consistent estimators for $\Var(X_{11})$ and $\Var(X_{21})$.
\bel{QnisQ}
\begin{split}
\Q^a(x_{1+}\frac{X_{+1}}{n}\frac{X_{+2}}{n}, x_{2+}\frac{X_{+1}}{n}\frac{X_{+2}}{n}) 
&= \frac{(X_{11} X_{22} - X_{12} X_{21})^2}{x_{2+}^2 x_{1+} \frac{X_{+1}}{n} \frac{X_{+2}}{n} + x_{1+}^2 x_{2+} \frac{X_{+1}}{n} \frac{X_{+2}}{n}}\\
&= \frac{(X_{11} X_{22} - X_{12} X_{21})^2}{x_{2+} x_{1+} X_{+1} X_{+2} \,\frac{1}{n^2} \,(x_{2+} + x_{1+})}\\
&= \frac{\big(X_{11}\,X_{22} - X_{12}\,X_{21}\big)^2}{x_{1+}\;x_{2+}\;X_{+1}\;X_{+2}\,\frac{1}{n}} \\
&= \Q.
\end{split}
\ee

\subsection{Derivation of \texorpdfstring{$\CMH^a (\Sighat_{1k}^2, \Sighat_{2k}^2)$}{}} \label{S_subsec_CMHa}

In \citet{agresti_categorical_2002} the test statistic is already presented in a form similar to \eqref{CMHStatA}. \\
The derivation can be carried out in a similar way as for the chi-square test. For each of the $\kstar$ partial 2x2 tables we obtain analogously to the chi-square test, compare \eqref{ConvVar1},
\bel{ConvVar1k}
\big(\frac{x_{2+k}}{n_k}\big)^2 \Sighat_{1k}^2 + \big(\frac{x_{1+k}}{n_k}\big)^2 \Sighat_{2k}^2 \toop \Var(\Ttildenk),
\ee 
with $\Ttildenk:= (X_{11k} - \frac{x_{1+k}\,X_{+1k}}{n_k})$.
Due to the independence of the partial 2x2 tables it is
\bel{VarSumIsSumVar}
\Var\Big(\sum_{k=1}^\kstar \Ttildenk \Big) = \sum_{k=1}^\kstar \Var(\Ttildenk).
\ee
Hence with the analogous considerations as carried out in section \ref{S_subsec_Qa} we obtain
\bel{Convcmh}
\CMH^a\big(\Sighat_{1k}^2, \Sighat_{2k}^2; k=1,\ldots,\kstar\big) \schw\chi^2_1,
\ee
with $\CMH^a\big(\Sighat_{1k}^2, \Sighat_{2k}^2; k=1,\ldots,\kstar\big)$ as defined in \eqref{CMHap}.

\subsection{Derivation of the variance estimators for the different scenarios} \label{S_subsec_ExplicitVars}

For the different scenarios we derive in the following the estimators of $\Var(X_{11})$ and $\Var(X_{21})$ for the chi-square test. The respective estimators for the CMH test can be deduced analogously.
\subsubsection{Genetic drift and one sampling step} \label{subsubsec_Drift}
We consider a bi-allelic SNP and assume to have allele counts from individual sequencing for allele 1 and allele 2 in the base population and a population that has evolved for $t$ generations. Table \ref{TabGen22Drift} is the corresponding contingency table and the sampling scheme is summarized in \mbox{figure \ref{SamplingSchemeOneSampling}}.

We assume that the null hypothesis of homogeneity is true.
Let be
\bel{NotationDrift}
X_{11}  \sim \Binom(x_{1+},p_1) \qquad \text{and} \qquad X_{21}  \sim \Binom(x_{2+},P_2),
\ee
corresponding to \eqref{AssBinDis}. Due to drift, the frequency of allele 1 after $t$ generations, $P_2$, is a random variable with \mbox{$\E[P_2] = p_1$}. This is in contrast to the original test, where $P_2=p_1$ under the null hypothesis. Note that still holds $E[T_n]=0$ as in equation \eqref{ETn0}.

Recall that we assume $X_{11}$ and $X_{21}$ to be independent. 
$p_1$ can consistently be estimated as $\frac{X_{11}}{x_{1+}}$. 
Drift affects only the evolved population, so we can simply infer from the Binomial distribution
\bel{VarX11Binom}
\Var(X_{11}) = x_{1+}\,p_1\,(1-p_1)
\ee
and
\bel{VarX11driftEst}
x_{1+}\,\frac{X_{11}}{x_{1+}}\,\frac{X_{12}}{x_{1+}} =  \frac{X_{11}\,X_{12}}{x_{1+}} \toop \Var(X_{11}).
\ee

For the calculation of $\Var(X_{21})$ we use the law of total variance:
\bel{VarX21drift}
\begin{split}
\Var(X_{21}) 
&= \E[\Var(X_{21}\,|\,P_2)] + \Var(\E[X_{21}\,|\,P_2])\\
&= \E[x_{2+}\,P_2\,(1-P_2)] + \Var(x_{2+}\,P_2)\\
&= x_{2+} \,\E[P_2]-x_{2+} \,\E[P_2^2] + x_{2+}^2 \Var(P_2)\\
&= x_{2+} \,\E[P_2]-x_{2+} \,\big(\Var(P_2) +\E[P_2]^2\big) + x_{2+}^2 \Var(P_2)\\
&= x_{2+} \Big(\E[P_2] \big(1 - \E[P_2]) + \big(x_{2+} - 1\big) \Var(P_2)\Big)
\end{split}
\ee

With the consistent estimators $\pbartwohat$ of $\E[P_2]$ and  $\sighatdriftsq$ of $\Var(P_2)$ we obtain
\bel{VarX21driftEstGen}
x_{2+} \Big(\pbartwohat \big(1 - \pbartwohat) + \big(x_{2+} - 1\big)\sighatdriftsq\Big) \toop \Var(X_{21}).
\ee

\subsubsection{No drift and two sampling steps} \label{subsubsec_Pool}

We now assume that all sequence data is from the same generation, and hence there is no drift, and we obtain the data from pool sequencing of a sample of the population. We model this as two binomial sampling steps.
For a given SNP we have  $r_1$ reads for the first and $r_2$ reads for the second population, i.e.\,$r_1$ and $r_2$ are the sequencing coverage values. 
Our model is to sample $r_1$ reads from a binomial distribution with probability  $\frac{X_{11}}{x_{1+}}$ for the first population and $r_2$ reads  with probability $\frac{X_{21}}{x_{2+}}$ for the second population. Table \ref{TabGen22Pool} is the corresponding contingency table.

Note that we only observe the data in table \ref{TabGen22Pool}, while we don't know the underlying data from table \ref{TabGen22Drift}, except for the numbers of sequenced individuals $x_{1+}$ and $x_{2+}$. 

Since we do two-step binomial sampling, we describe first a general two-step binomial sampling scenario. 
In \citep{jonas_estimating_2016} similar calculations are carried out.
Let be
\bel{Gen2StepBinSam}
K_1 \sim \Binom(g_1, b_0) \qquad \text{and} \qquad K_2 \sim \Binom(g_2, B_1) \quad \text{with } B_1=\frac{K_1}{g_1}.
\ee
With the law of total variance we obtain

\begin{align}
\Var(K_2)
&= \E[\Var(K_2\,|\,K_1)] + \Var(\E[K_2\,|\,K_1]) \nonumber \\
&= \frac{g_2}{g_1}\, \Big(\E[K_1]  \Big(1 - \frac{\E[K_1]}{g_1} \Big) + \frac{g_2-1}{g_1} \Var(K_1) \Big)  \label{VarGen2StepBinSam1} \\
&= g_2 \, b_0 \,(1-b_0) \Big(1 + \frac{g_2-1}{g_1}\Big). \label{VarGen2StepBinSam2}
\end{align}

$\bigxhat_{11}$ corresponds to $K_2$.
In the first step we sample $x_{1+}$ alleles, and the probability for allele 1 is $p_1$.
The second step is the sampling of the read counts as described above. 
With \eqref{VarGen2StepBinSam2} and estimating $p_1$ as $\frac{\bigxhat_{11}}{r_1}$ we can set
\bel{EstVar1Pool}
\frac{\bigxhat_{11}\,\bigxhat_{12}}{r_1}\Big(1 + \frac{r_1-1}{x_{1+}}\Big) \toop \Var(\bigxhat_{11}).
\ee

Note that without drift, we have $X_{21}  \sim \Binom(x_{2+},p_2)$, where $p_2$ is not a random variable.
For population 2 we have hence analogously
\bel{EstVar2Pool}
\frac{\bigxhat_{21}\,\bigxhat_{22}}{r_2}\Big(1 + \frac{r_2-1}{x_{2+}}\Big) \toop \Var(\bigxhat_{21}).
\ee

\subsubsection{Genetic drift and two sampling steps} \label{subsubsec_DriftPool}

We now consider a scenario where additional variance is induced by drift as well as pool sequencing. 
We model drift and pool sequencing as previously described, the sampling scheme is summarized in figure \ref{SamplingSchemeTwoSampling}. 
As drift does not affect the base population, we can estimate $\Var(\bigxhat_{11})$ again as in \eqref{EstVar1Pool}.

For the estimator of $\Var(\bigxhat_{21})$ we use equation \eqref{VarGen2StepBinSam1}. Since $P_2$ is a random variable like in \eqref{NotationDrift}, the variance of $X_{21}$ is different from the binomial sampling variance and \eqref{VarGen2StepBinSam2} does not hold here.  Similar calculations can be found in \citep{jonas_estimating_2016}.

Further, with the law of total expectation it is
\bel{EX21Dr}
\E[X_{21}] = \E[\E[X_{21}\,|\,P_2]] = x_{2+}\,\E[P_2].
\ee
With \eqref{VarGen2StepBinSam1}, \eqref{VarX21drift}, and \eqref{EX21Dr} we obtain 
\bel{VarEvolDrPool}
\begin{split}
\Var(\bigxhat_{21}) 
&= \frac{r_2}{x_{2+}}\, \Big(\E[X_{21}]  \Big(1 - \frac{\E[X_{21}]}{x_{2+}} \Big) + \frac{r_2-1}{x_{2+}} \Var(X_{21}) \Big) \\
&= r_2\Big(\E[p_2]  (1-\E[p_2]) \big(1 + \frac{r_2-1}{x_{2+}}\big) + (r_2-1)\,\frac{x_{2+}-1}{x_{2+}} \Var(P_2)\Big).
\end{split}
\ee

With the consistent estimators $\pbartwohat$ of $\E[P_2]$ and  $\sighatdriftsq$ of $\Var(P_2)$ we obtain then 
\bel{VarX21driftpoolEstGen}
r_2\Big(\pbartwohat  (1-\pbartwohat) \big(1 + \frac{r_2-1}{x_{2+}}\big) + (r_2-1)\,\frac{x_{2+}-1}{x_{2+}} \sighatdriftsq \Big) \toop \Var(\bigxhat_{21}).
\ee

\subsection{P-value Correction } \label{S_correc_P}
\begin{figure}[H]
\subfigure[correction function
  \label{fig:transformation}]{\includegraphics[width=0.49\textwidth]{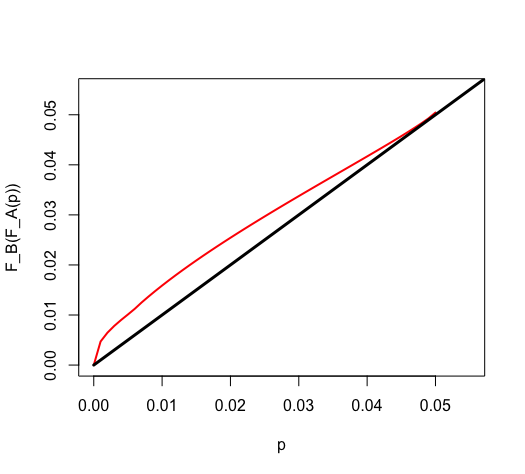}}\hfill
\subfigure[correction result
  \label{fig:result}]{\includegraphics[width=0.49\textwidth]{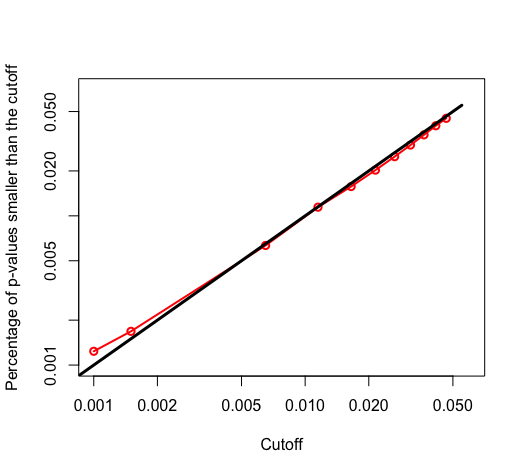}} 
\caption{(a) Function $F_t(\cdot)$ used to transform small p-values.
(b) Percentage of loci with p-value smaller than cut-off value against cut-off value for $T^{2s\&d}_{\chi^2}$. The black line is the angle bisector. Simulation setup: $10^6$ neutral loci with true allele frequencies in base population uniformly distributed on $[0, 1]$, $Ne = 300$, allele sample size 1000, pool sequencing with coverage Poisson distributed with mean 80, sequence data for generations 0 and 60.}
\label{TrafoRes}
\end{figure}

\begin{figure}[h]
\subfigure[Effective population size $N_e$]{\includegraphics[width=0.423\textwidth]{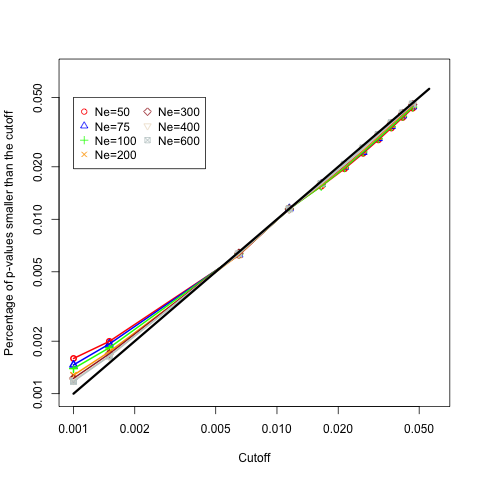}}\hfill
\subfigure[Pool size]{\includegraphics[width=0.423\textwidth]{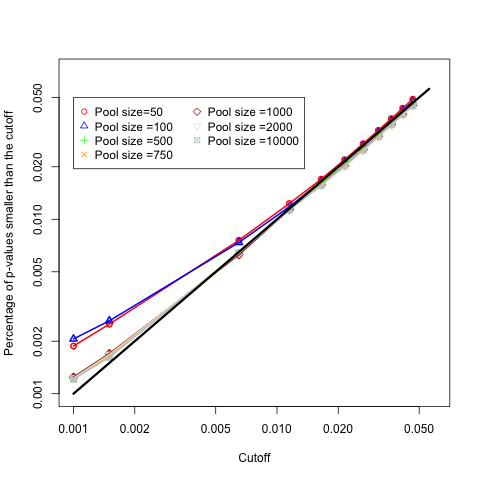}} \\
\subfigure[Coverage]{\includegraphics[width=0.423\textwidth]{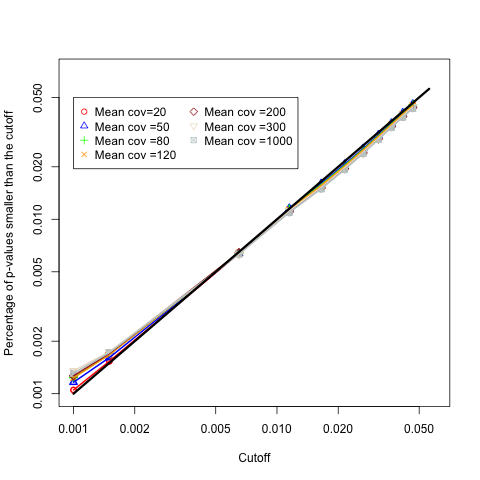}}\hfill
\subfigure[Number of generations]{\includegraphics[width=0.423\textwidth]{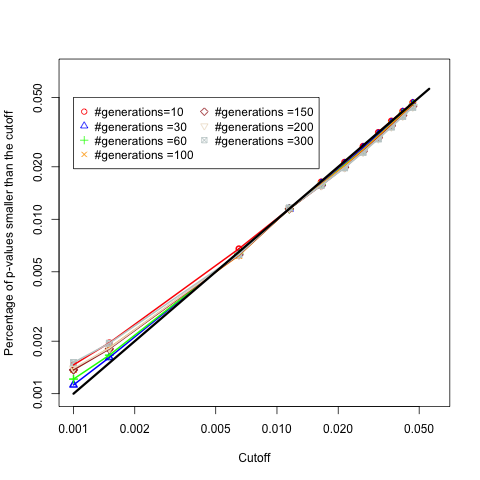}} \\
\subfigure[Starting allele frequency]{\includegraphics[width=0.423\textwidth]{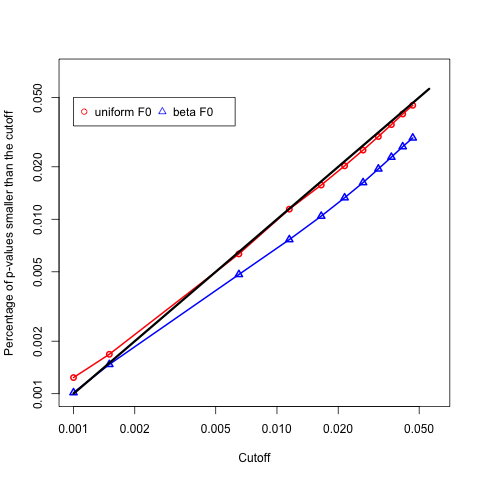}}\hfill
\subfigure[Allele frequency range]{\includegraphics[width=0.423\textwidth]{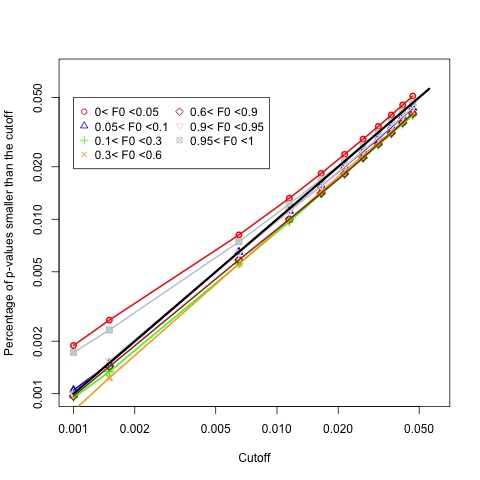}} \\
\caption{Percentage of loci with p-value smaller than cut-off value against cut-off value for $T^{2s\&d}_{\chi^2}$ with the correction for the significant p-values. The black line is the angle bisector. Simulation setup: For each plot one parameters varies according to the legend and the others follow the following setup. $10^6$ neutral loci with true allele frequencies in base population uniformly distributed on $[0, 1]$, $Ne = 300$, allele sample size 1000, pool sequencing with coverage Poisson distributed with mean 80, sequence data for generations 0 and 60.}
\label{fig:parameters}
\end{figure}

To obtain a more uniform tail distribution for the null p-values obtained with the chi-square test, we propose a correction that uses
the following functions:
\begin{itemize}
    \item $F_A(x)$ and $F_{A2}(x)$ are beta cumulative distribution functions with suitably chosen parameters
    \item $F_B(x) := \begin{cases} s F_{A2}(x/s) & 0 \leq x < s\\
\delta + \frac{x - \delta}{1 - \delta} & s \leq x \leq 1\end{cases}$ \\ where 
$\delta = s F_{A2}(1)$, and suitably chosen $s \in [0, z]$, $z$ in $[0,1]$. 
\end{itemize}
For a  p-value $\pi_z$, smaller than the chosen threshold $z$, we transform $ \pi^\ast_z := F_t(\pi_z):=  F_B(z\cdot F_A(\pi_z/z))$.
The parameters of the beta distributions have been obtained using method of moments estimates for the rescaled data, i.e. $\pi_z/z$ 
for $F_A(\cdot)$, and $\pi_s/s$ with $\pi_s=z\cdot F_A(\pi_z/z))$ for $F_{A2}(\cdot).$ 
Figure \ref{fig:transformation} displays the transformation $F_t(\cdot)$, and in figure \ref{fig:result} the improvement may be seen
achieved by the transformation.
Figure \ref{fig:parameters} suggests that some models and circumstances require different transformations.

\subsection{Additional figures} \label{S_subsec_SupFigures}
\hspace{2cm}

\begin{figure}[h]
\subfigure[$\Q$
  \label{HistUnifQmClas}]{\includegraphics[width=0.49\textwidth]{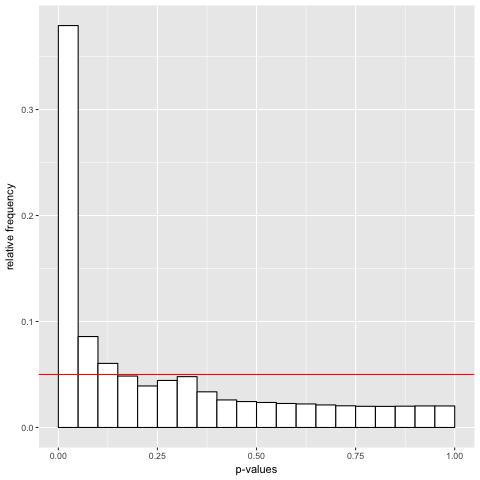}}\hfill
\subfigure[$\Qmdrpoolp$
  \label{HistUnifCmhA1}]{\includegraphics[width=0.49\textwidth]{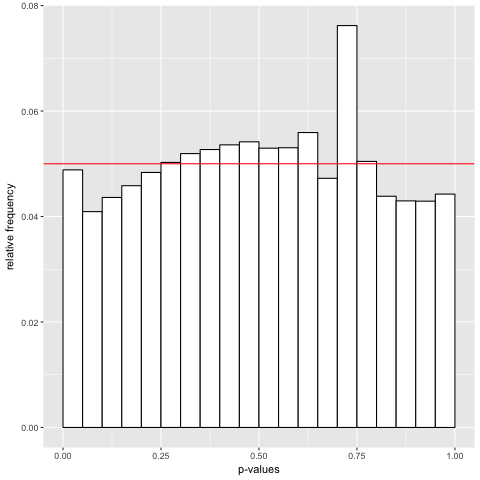}} 
\caption{Distribution of the p-values belonging to the classical chi-square test \mbox{$\Q$ (a)} and $\Qmdrpoolp$ (b). Red line indicates frequency of $5\%$. Simulation setup: $10^6$ neutral loci with true allele frequencies in base population uniformly distributed on $[0,1]$, $N_e=300$, allele sample size 1000, pool sequencing with coverage Poisson distributed with mean 80, sequence data for generations 0 and 60.}
\label{HistUnifQmClasA}
\end{figure}

\begin{figure}[h]
\subfigure[$\Qndrp$
  \label{HistUnifQmD}]{\includegraphics[width=0.49\textwidth]{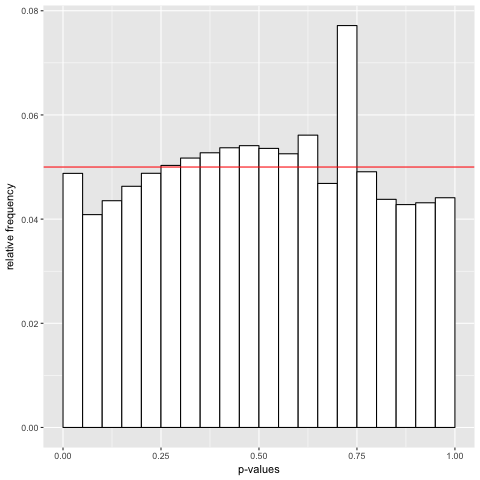}}\hfill
\subfigure[$\CMHdrp$
  \label{HistUnifCmhD}]{\includegraphics[width=0.49\textwidth]{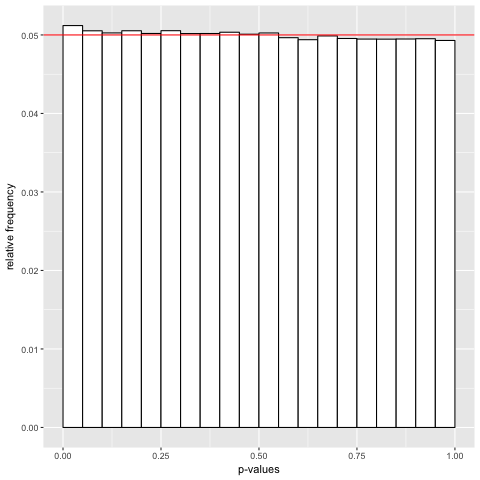}} 
\caption{Distribution of the p-values belonging to $\Qndrp$ (a) and $\CMHdrp$ (b). Red line indicates frequency of $5\%$. Simulation setup: $10^6$ neutral loci with true allele frequencies in base population uniformly distributed on $[0,1]$, $N_e=300$, pool sequencing of whole population with coverage Poisson distributed with mean 80, sequence data for generations 0 and 60, 3 replicate populations in (b).}
\label{HistUnifDrift}
\end{figure}

\begin{figure}[h]
\subfigure[$\Q$
  \label{HistUnifQmPClas}]{\includegraphics[width=0.49\textwidth]{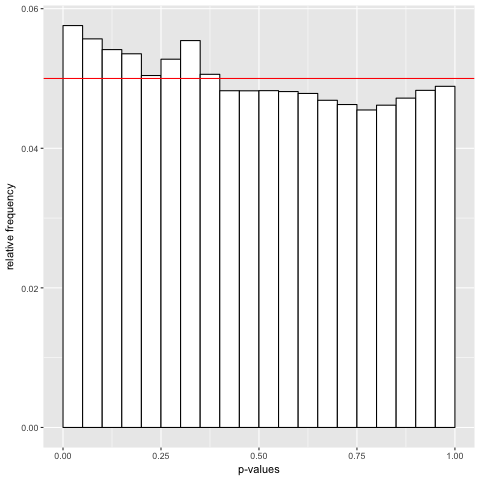}}\hfill
  \subfigure[chi-square test adapted to 2 sampling steps
  \label{HistUnifQmP}]{\includegraphics[width=0.49\textwidth]{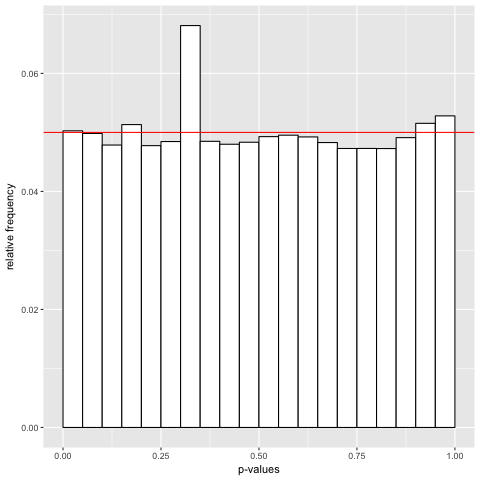}}\hfill
\subfigure[$\CMH$
  \label{HistUnifCmhPClas}]{\includegraphics[width=0.49\textwidth]{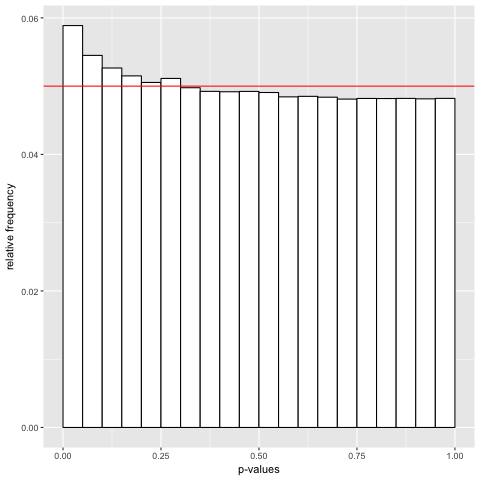}} 
  \subfigure[CMH test adapted to 2 sampling steps
  \label{HistUnifCmhP}]{\includegraphics[width=0.49\textwidth]{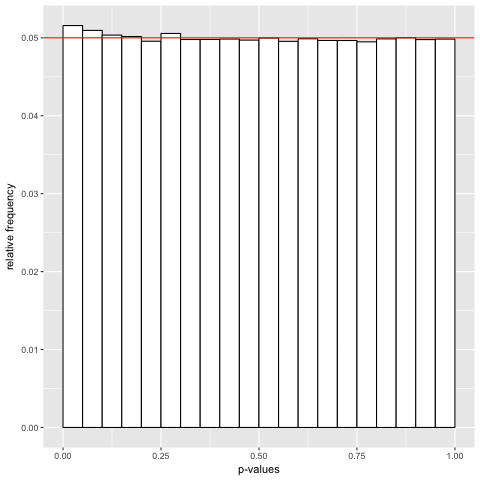}} 
\caption{Distribution of the p-values belonging to the classical chi-square test  $\Q$ (a), the chi-square test adapted to 2 sampling steps (b), the classical CMH test $\CMH$ (c), the CMH test adapted to 2 sampling steps (d). Red line indicates frequency of $5\%$. Simulation setup: $10^6$ neutral loci with true allele frequencies in population 1 uniformly distributed on $[0,1]$ and the same true frequencies in population 2 (no drift), allele sample size 1000, pool sequencing with coverage Poisson distributed with mean 80, 3 replicate populations in (c) and (d).}
\label{HistUnifPool}
\end{figure}

\begin{figure}[h]
\subfigure[$\Qmdrpoolintgenp$
  \label{PowerNointgenVsIntgenQm}]{\includegraphics[width=0.49\textwidth]{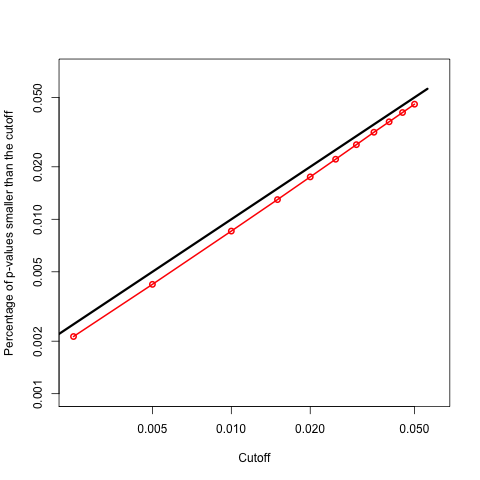}}\hfill
\subfigure[$\CMHdrpoolintgenp$
  \label{PowerNointgenVsIntgenCMH}]{\includegraphics[width=0.49\textwidth]{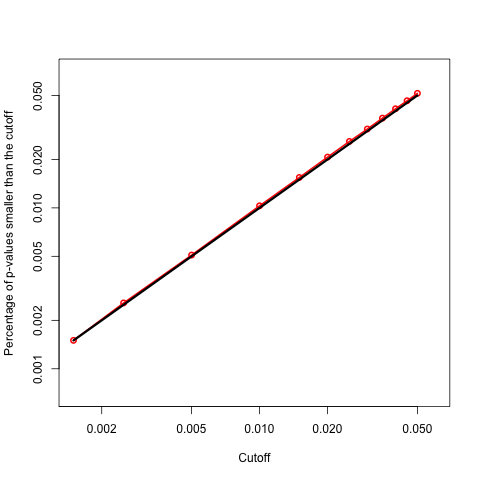}} 
\caption{Percentage of loci with p-value smaller than cut-off value against cut-off value in (a) for $\Qmdrpoolintgenp$, and in (b) for $\CMHdrpoolintgenp$. The black line indicates the angle bisector. Simulation setup: $10^6$ neutral loci with true allele frequencies in base population uniformly distributed on [0,1], Ne = 300, allele sample size 1000, pool sequencing with coverage Poisson distributed with mean 80, sequence data for generations 0, 10, 20, 30, 40, 50, 60; 5 replicate populations in (b).}
\label{ConservativenessIntgen}
\end{figure}

\begin{figure}[h]
\subfigure[20 generations of drift
  \label{PowerDepSCMH20}]{\includegraphics[width=0.49\textwidth]{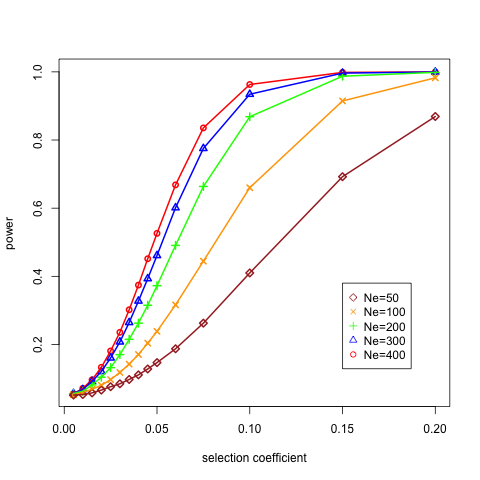}}
\subfigure[200 generations of drift
  \label{PowerDepSCMH200}]{\includegraphics[width=0.49\textwidth]{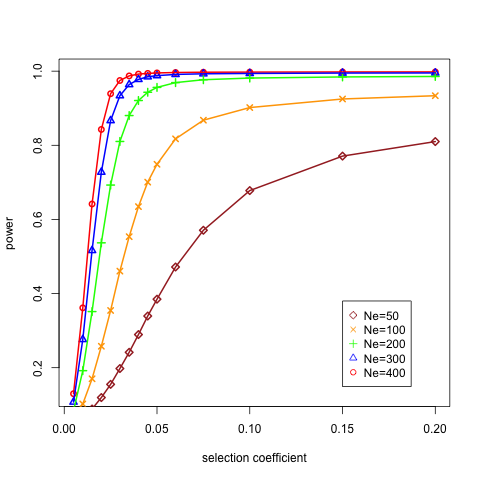}}
\caption{Power of $\CMHdrpoolp$ vs selection coefficients for different values of the effective population size $N_e$. Simulation setup: $10^5$ loci for each selection coefficient with true allele frequencies in base population uniformly distributed on $[0,1]$, allele sample size 1000, pool sequencing with coverage 100, 5 replicate populations. (a) sequence data generations 0 and 20, (b) sequence data generations 0 and 200.}
\label{PowerDepS_supCMH}
\end{figure}

\begin{figure}[h]
\subfigure[$\Qmdrpoolp$
  \label{PowerDepSQmBeta}]{\includegraphics[width=0.49\textwidth]{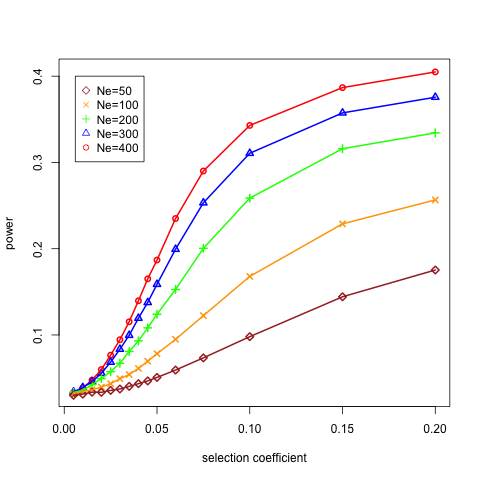}}
\subfigure[$\CMHdrpoolp$
  \label{PowerDepSCMHBeta}]{\includegraphics[width=0.49\textwidth]{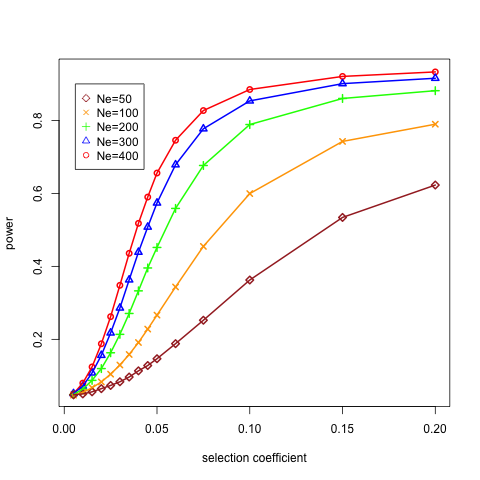}}
\caption{Power of $\Qmdrpoolp$ (a) and $\CMHdrpoolp$ (b) vs selection coefficients for different values of the effective population size $N_e$. Simulation setup: $10^5$ loci for each selection coefficient with true allele frequencies in base population beta distributed with parameters 0.2 and 0.2, allele sample size 1000, pool sequencing with coverage 100, sequence data for generations 0 and 60, 5 replicate populations in (b).}
\label{BetaPower}
\end{figure}

\end{document}